\documentclass[twocolumn,iop,apj]{aastex63}
\usepackage{amsfonts,amsmath,graphicx,natbib,url,hyperref,nicefrac}
\usepackage{amssymb, verbatim, subfigure, paralist, soul, comment, multirow}

\hypersetup{colorlinks=true, urlcolor=blue, citecolor=blue}

\usepackage{color}
\newcommand{\mesa}{\texttt{MESA\;}}
\newcommand{\sprout}{\texttt{Sprout\;}}
\newcommand{\stella}{\texttt{Stella\;}}

\shorttitle{Core-collapse SNe in binaries} 
\shortauthors{Mandal et al.}

\begin{document}

\title{Same explosion, many faces: numerical modeling reveals viewing angle as a driver of diversity for core-collapse SNe in binary systems}

\author[0000-0001-9484-1262]{Soham Mandal}
\affiliation{Department of Astronomy, University of Virginia, 530 McCormick Road, Charlottesville, VA 22904, USA}
\affiliation{Virginia Institute for Theoretical Astronomy, University of Virginia, Charlottesville, VA 22904, USA}

\author[0009-0004-7268-7283]{Raphael Baer-way}
\affiliation{Department of Astronomy, University of Virginia, 530 McCormick Road, Charlottesville, VA 22904, USA}
\affiliation{National Radio Astronomy Observatory,
520 Edgemont Rd, Charlottesville VA 22903, USA}

\author[0000-0002-1856-9225]{Shazrene Mohamed}
\affiliation{Department of Astronomy, University of Virginia, 530 McCormick Road, Charlottesville, VA 22904, USA}
\affiliation{Virginia Institute for Theoretical Astronomy, University of Virginia, Charlottesville, VA 22904, USA}
\affiliation{South African Astronomical Observatory, P.O Box 9, Observatory, 7935, Cape Town, South Africa}
\affiliation{Department of Astronomy, University of Cape Town, Private Bag X3, Rondebosch, 7701, Cape Town, South Africa}
\affiliation{NITheCS National Institute for Theoretical and Computational Sciences, South Africa}

\author[0000-0003-1450-0869]{Irene Salmaso}
\affiliation{INAF–Osservatorio Astronomico di Capodimonte, Salita Moiariello 16, 80131 Napoli, Italy}

\author[0000-0002-0844-6563]{Poonam Chandra}
\affiliation{National Radio Astronomy Observatory,
520 Edgemont Rd, Charlottesville VA 22903, USA}
\affiliation{Department of Astronomy, University of Virginia, 530 McCormick Road, Charlottesville, VA 22904, USA}

\author[0000-0001-7132-0333]{Maryam Modjaz}
\affiliation{Department of Astronomy, University of Virginia, 530 McCormick Road, Charlottesville, VA 22904, USA}

\email{soham@virginia.edu}

\begin{abstract}

Observable properties of core-collapse supernovae (CCSNe) depend sensitively on the circumstellar material (CSM) formed by pre-explosion mass loss from the progenitor star. Since a large fraction of CCSN progenitors reside in binaries, both the progenitor structure and surrounding CSM can be significantly impacted by binary interaction. Yet, its impact on the observed CCSN landscape remains poorly constrained. In this work, we investigate CCSNe from binary systems undergoing stable Roche lobe overflow. We construct a suite of binary evolution models in \texttt{MESA} with a fixed initial primary mass ($16M_{\odot}$), exploring secondary masses in the range $12-15M_{\odot}$ and initial orbital periods $>500$ days. We generate three-dimensional CSM structures from the resulting mass-loss histories and orbital dynamics, extract angle-dependent density profiles along three lines of sight, and compute multi-band light curves with the radiation-hydrodynamics code \texttt{Stella}. We find that binary-driven CSM develops highly aspherical morphologies, governed by the orbital period and the mass ratio. Interaction between SN ejecta and this structured medium produces pronounced viewing-angle dependence in the light curves, with peak luminosities differing by a factor of $\sim5$ and late-time $B-V$ colors varying by $\sim1.5$ mag depending on observer orientation. We further show that interpreting such events with one-dimensional frameworks assuming isolated progenitors and spherical winds can introduce biases up to $50\%$ for inferred explosion properties and $>200\%$ for inferred mass-loss rates. Our results are consistent with a substantial fraction of interacting Type II SN diversity arising from binary-shaped asymmetric CSM and viewing-angle effects, motivating multidimensional approaches to interpreting these transients.

\end{abstract}


\section{Introduction}  \label{sec:intro}

Massive stars below $\sim 150M_{\odot}$ likely end their lives as core-collapse supernovae (CCSNe), injecting energy, momentum, and newly synthesized elements into the interstellar medium and thereby playing a central role in galactic evolution \citep[e.g.,][]{Smartt2009ARAandA,Janka+2012PTEP,Woosley+2015ApJ}. Observable signatures of CCSNe are shaped not only by the explosion physics and progenitor structure, but also by the circumstellar material (CSM) into which the ejecta expand \citep[for reviews, see e.g.,][and references therein]{Smith2014ARAandA, Chandra2018, Fraser2020RSOS, Margalit2022}. While early studies on massive star evolution largely focused on isolated stars, it is now well established that a large fraction of massive stars \citep[$\sim71\%$ of O-stars; e.g.,][]{Sana+2012Sci} reside in multiple systems, including binaries, and undergo interaction during their lifetimes \citep{Duchene2013ARAandA,Moe+2017ApJS}. Binary interaction introduces physical processes absent in single star evolution, such as Roche lobe overflow \citep[RLOF;][]{Lubow+1975ApJ}, common envelope evolution \citep[CEE;][]{Chevalier2012ApJ}, tidal interactions and stellar mergers \citep{Podsiadlowski1992ApJ,Tauris+2006book,deMink+2013ApJ}. These processes can alter the structure and composition of the exploding star. In particular, mass transfer can strip the hydrogen-rich envelope. This process produces progenitors of stripped-envelope supernovae or SESNe \citep[Types IIb, Ib, and Ic;][]{Filippenko1997ARAandA,Modjaz+2019NatAs}, in contrast to the hydrogen-rich Type IIP/L SNe expected from red supergiants or RSGs that have not experienced significant stripping \citep{Yoon+2010ApJ,Eldridge+2013MNRAS,Dessart+2015MNRAS}. More generally, binary interactions can modify progenitor properties throughout their evolution \citep{Langer2012ARAandA,Laplace+2021AandA,Schneider+2024AandA,Schneider+2025arXiv}. At the same time, they can redistribute substantial amounts of mass and angular momentum into the surrounding environment through mass transfer, outflows, and envelope ejection. Consequently, binary interactions have the potential to shape both the progenitor and the circumstellar medium in which the eventual SN explosion occurs.

\begin{figure*}
\vspace{2mm}
\centering
\includegraphics[width=0.98\textwidth]{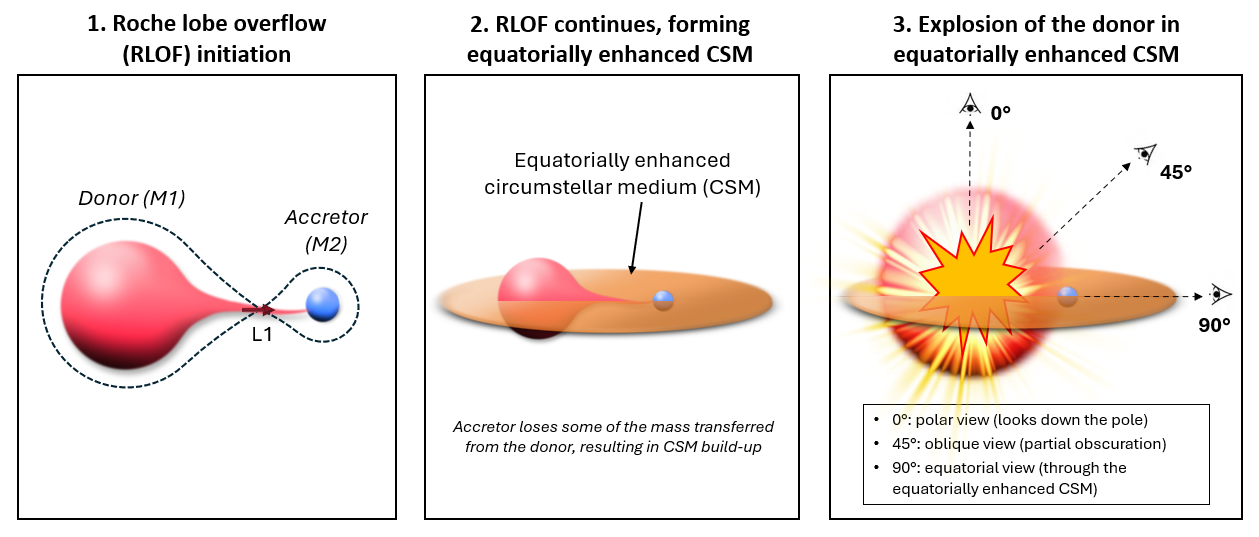}
\caption{Cartoon illustration (not to scale) of the physical scenario considered in this work. Stable, non-conservative Roche lobe overflow (RLOF) from the donor star produces an equatorially enhanced circumstellar medium (CSM). The donor subsequently undergoes core-collapse, and the resulting SN ejecta interacts with the asymmetric CSM. Angle-dependent light curves are obtained by considering representative viewing angles of $\theta=0^{\circ},45^{\circ},\mathrm{\,and\,}90^{\circ}$ relative to the symmetry axis. The stellar radii, compared to the extent of the CSM, have been highly exaggerated for the sake of clarity.}
\label{fig:cartoon}
\end{figure*}

%uch as the Palomar Transient Factory \citep[PTF;][]{Law+2009PASP}, the All-Sky Automated Survey for SuperNovae \citep[ASAS-SN;][]{Kochanek+2017PASP}, the Zwicky Transient Facility (ZTF) Bright Transient Survey \citep{Perley+2020ApJ}, the Young Supernova Experiment \citep[YSE;][]{Aleo+2023ApJS}, and the upcoming Legacy Survey of Space and Time \citep[LSST;][]

Evidence for complex circumstellar environments around CCSNe has grown rapidly with the advent of modern high cadence surveys such as the Palomar Transient Factory \citep[PTF;][]{Law+2009PASP}, the All-Sky Automated Survey for SuperNovae \citep[ASAS-SN;][]{Kochanek+2017PASP}, the Zwicky Transient Facility (ZTF) Bright Transient Survey \citep{Perley+2020ApJ}, the Young Supernova Experiment \citep[YSE;][]{Aleo+2023ApJS}, and the Legacy Survey of Space and Time \citep[LSST;][]{Ivezic+2019ApJ}. Early-time spectroscopy has revealed ``flash-ionization'' features produced by dense material confined close ($\sim10^{14}$ cm) to the progenitor \citep{GY+2014Nature,Bruch+2021ApJ}. Recent well-observed events such as SN 2023ixf have further highlighted the presence of nearby dense material around Type II SNe \citep{JG+2023ApJ,Smith+2023ApJ}. In addition, $\approx10\%$ of CCSNe show optical signatures of significant CSM interaction \citep{Perley+2020ApJ}. In such strongly interacting events (Types IIn, Ibn, and Icn), ejecta-CSM shocks can convert a substantial fraction of the kinetic energy into radiation \citep[up to $\sim10\%$, compared to $\sim1\%$ for typical CCSNe; see][]{Dessart2024arXiv}. These observations demonstrate that the final stages of massive-star evolution can produce a wide range of circumstellar environments. Moreover, understanding CCSN diversity may require characterizing not only the amount of CSM, but also its geometry. Indeed, asymmetric CSM has been inferred in several systems through detailed multiwavelength photometry, spectroscopy \citep{Chandra+2020ApJ,BW+2025ApJ}, and spectropolarimetry \citep{Bilinski2024}.

The diversity of CSM structures is difficult to reconcile with steady line-driven winds \citep{Castor+1975ApJ,Puls+2008AandARv} or dust-driven winds \citep{vanLoon2000AandA,Goldman+2017MNRAS} from massive stars \citep{Smith2014ARAandA,Smith2017hsn..book}. Many channels of complex CSM structure formation have been proposed, such as pulsational instabilities \citep{Woosley+2002RvMP,Woosley+2007Natur,Ma+2025arXiv,Suzuki2025,Sengupta2026}, turbulent convection induced unsteady nuclear burning \citep{Smith+2014ApJ}, wave-driven mass loss \citep{Quataert+2012MNRAS,Shiode+2014ApJ}, and binary interactions. While each of these mechanisms can generate complex circumstellar environments, binary interactions are unusual in their ability to modify both the progenitor star and its surroundings over evolutionary timescales extending well before core collapse.

The dual role of binary interaction in shaping both progenitors and circumstellar environments has motivated extensive theoretical work. Detailed grids of binary stellar evolution models have explored how binary interactions modify the internal structure and mass-loss histories of CCSN progenitors \citep{Zapartas+2021AandA,Ercolino+2024AandA,Gilkis+2025MNRAS,Tsai+2026arXiv}. Three-dimensional (3D) hydrodynamic simulations have investigated the formation of complex, asymmetric CSM structures through RLOF, CEE, and wind mass transfer \citep{Mohamed+2012BaltA,Pejcha+2016MNRAS,MacLeod+2020ApJ_a}. Multidimensional radiation-hydrodynamic simulations have further demonstrated that SNe exploding into asymmetric environments can exhibit strongly viewing-angle-dependent observational signatures \citep{Vlasis+2016MNRAS,Suzuki+2019ApJ}. Despite this progress, connecting binary evolution, multidimensional CSM formation, and observable signatures of CCSNe affected by binary interaction remains challenging because these processes span many orders of magnitude in spatial and temporal scale.

In this work, we take a step toward addressing this problem. We construct a suite of one-dimensional (1D) binary evolution models spanning a range of initial masses and orbital configurations, focusing on regimes of stable RLOF. Using the resulting mass-loss histories and orbital properties, we perform 3D hydrodynamic computations to characterize the circumstellar environments produced by these systems. We then extract angle-dependent density profiles from the 3D models and compute multi-band SN light curves using 1D radiation-hydrodynamic calculations along multiple lines of sight. \autoref{fig:cartoon} provides a schematic overview of the physical processes modeled in this work, from the onset of RLOF and CSM formation to the subsequent SN and the viewing-angle dependent SN emission. Our goal is to investigate how variations in binary properties and viewing angle shape observed SN light curves and to identify regimes in which binary-generated circumstellar structure can significantly affect CCSN observables.

The modeling approach and techniques are given in \autoref{sec:numerical}. We present our results in \autoref{sec:results}, followed by a discussion in \autoref{sec:discussion}, and the conclusions in \autoref{sec:conclusion}.

\begin{figure*}
\vspace{2mm}
\centering
\includegraphics[width=0.98\textwidth]{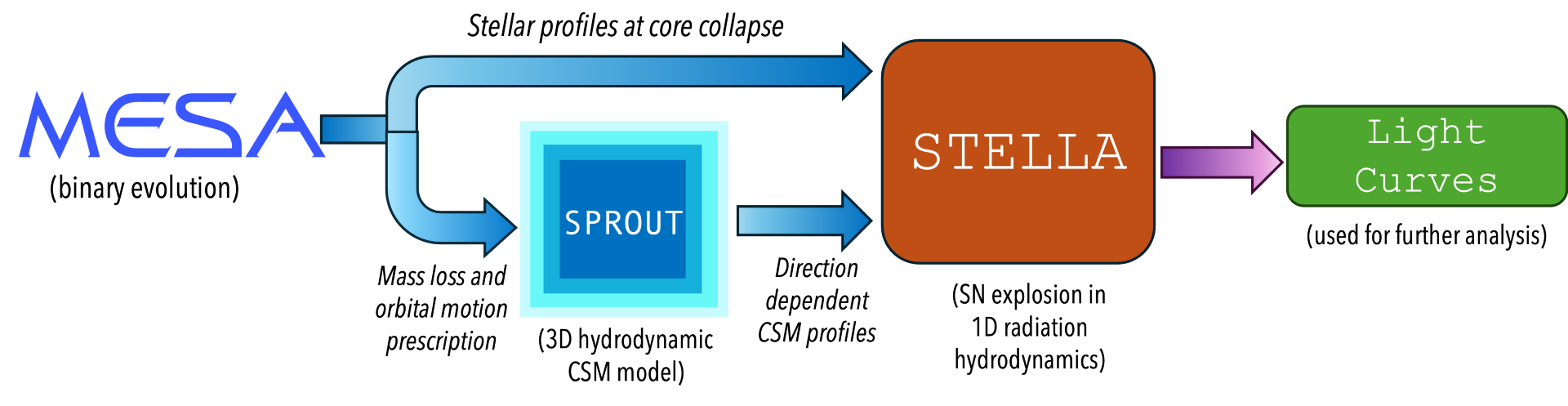}
\caption{Schematic diagram of the method used here. Binary stellar models are evolved in \mesa up to the core collapse of the initially more massive star. The resulting mass-loss rates from the binary, as well as the orbital dynamics, are used as initial conditions in the hydrodynamics code \sprout to compute 3D CSM models. Direction-dependent 1D profiles are extracted from the \sprout models and appended to the core collapse models from \mesa. The resulting explosion of the star into the different direction-dependent 1D CSM profiles is computed using \stella. This produces multi-band light curves of the SN explosion, as would be viewed from different viewing angles. These light curves are further analyzed using the technique detailed in Section~\ref{subsec:fitting}.}
\label{fig:schematic}
\end{figure*}

\section{Numerical method} \label{sec:numerical}

We develop a set of models for wide massive binaries using the 1D stellar evolution code \mesa \citep[r24.08.1;][]{Paxton2011, Paxton2013, Paxton2015, Paxton2018, Paxton2019, Jermyn2023}, in which both stars are evolved simultaneously. In these systems, non-conservative mass transfer via RLOF results in the formation of CSM around the binary, which is modeled using the 3D hydrodynamic code \texttt{Sprout}. Upon explosion of the donor star, the stellar ejecta expands against this CSM. The resulting asymmetric ejecta-CSM interaction is modeled with multiple 1D radiation hydrodynamic calculations using the code \stella \citep{Blinnikov+1993AandA,Blinnikov+1998ApJ,Blinnikov+2004ApandSS,Blinnikov+2006AandA} per stellar model, each corresponding to a hydrodynamic profile extracted along a distinct viewing angle from the 3D CSM models. A schematic of our modeling pipeline is provided in \autoref{fig:schematic}.

By performing multiple 1D calculations rather than a single multi-D model per system, we incorporate frequency-dependent opacities and realistic thermodynamics that are challenging to include at comparable fidelity in fully multi-D treatments, which must balance dimensionality with radiation transport complexity \citep{Suzuki+2019ApJ,Kurfurst+2026arXiv}. At the same time, dominant impacts of the asymmetric CSM are captured using angle-dependent profiles extracted from the 3D CSM models. This framework therefore enables a systematic exploration of direction-dependent observables from SNe in binary systems, even beyond the parameter space considered here.

\subsection{Binary modeling using \mesa} \label{subsec:mesa}

The evolution of massive stars in binaries depends sensitively on both the orbital period and the stellar masses \citep{Yoon+2017ApJ,Ercolino+2024AandA}. The extensive multi-D parameter space of binary stellar evolution requires systematic exploration to robustly connect binary progenitors to CCSN outcomes \citep{Gilkis+2025MNRAS,Jin+2026AandA}. In this exploratory study, we consider a small representative subset of this parameter space. We model systems with a primary star (hereafter the donor) of initial mass ($M_{d,i} = 16M_{\odot}$). This star is in orbit with a companion (which is initially less massive and later becomes the accretor), with a period of $P_i$ days. The initial mass of the companion is set as $M_{a,i}=q_i M_{d,i}$, where $q_i$ is the initial mass ratio \citep[following the convention of][]{Ercolino+2024AandA}. We evolve a set of models spanning selected combinations of ($P_i$, $q_i$), whose details are summarized in Table~\ref{tab:mesa_models}.

The chosen systems are constructed such that RLOF is initiated either during core helium burning (Case B) or after helium depletion in the core (Case C). In these regimes, the donor undergoes substantial mass stripping \citep{Tauris+2006book}, significantly altering its pre-SN structure and facilitating the formation of stripped-envelope or interacting SNe. Moreover, mass transfer and loss during Case B/C RLOF can lead to the buildup of dense CSM in the vicinity of the system and subsequent SN–CSM interaction \citep{Matsuoka+2024ApJ}. We restrict our study to systems undergoing stable mass transfer and do not model common envelope evolution (CEE), where mass transfer becomes dynamically unstable.

\subsubsection{Metallicity, opacities, and nuclear network}

All models are assumed to have solar metallicity \citep[Z=0.0154;][]{Asplund+2021AandA}, Z being the total mass fraction of all elements heavier than helium.

Radiative opacity values were primarily adopted from OPAL \citep{Iglesias1993,Iglesias1996}, with low-temperature ($\mathrm{log\,T[K]}<4$) data by \citet{Ferguson2005} and the high-temperature regime by \citet{Poutanen2017}. Electron conduction opacities are adopted from \citet{Cassisi2007} and \citet{Blouin2020}.

We employ the $\texttt{approx21\_cr60\_plus\_co56.net}$ reaction network available in \mesa. Reaction rates are adopted from the JINA REACLIB database \citep{Cyburt2010}, NACRE \citep{Angulo1999} and additional tabulated weak reaction rates are from \citet{Fuller1985,Oda1994,Langanke2000}. Screening is accounted for using the prescription of \citet{Chugunov2007}. Thermal neutrino-loss rates are from \citet{Itoh1996}.

\subsubsection{Mixing}

Mixing in the convective layers of the star is calculated according to the Time Dependent Convection (TDC) formulation in \mesa\citep{Jermyn2023}, with the mixing length parameter $\alpha_{\mathrm{MLT}}=1.5$. Superadiabatic regions are assumed to be convective if they satisfy the Ledoux criterion, and semiconvective otherwise \citep{Langer+1985AandA} with an efficiency parameter $\alpha_{\mathrm{sc}}=1.0$ \citep{Yoon+2010ApJ}. Convective overshooting of the hydrogen burning core and the helium burning core is implemented following \cite{Paxton2013}. The overshooting length is chosen to be $0.035H_p$ and $0.01H_p$ respectively, where $H_p$ is the pressure scale height at the boundary of the convective core. We include thermohaline mixing with an efficiency parameter $\alpha_{\mathrm{th}}=2.0$ \citep[for details, see][]{Paxton2015}. Rotational mixing is assumed to occur in these models, with a rotational diffusion coefficient multiplier of $0.033$ \citep{Heger+2000ApJ}.

\subsubsection{Stellar wind}

Mass loss due to stellar winds is calculated using the ``Dutch" prescription in \mesa, with a scaling factor of $1.0$. This wind scheme follows that of \cite{deJager+1988AandAS} for cool stars ($\mathrm{T_{\rm eff}}<10^4\rm K$). This prescription has been validated through observations of galactic red supergiants \citep[RSGS;][]{Mauron+2011AandA}, although recent studies of RSGs in clusters suggests that this prescription may be overestimating the true mass loss by a factor of a few tens \citep{Beasor+2020MNRAS,Beasor+2021ApJ}. Nevertheless, as will be shown in Section~\ref{subsec:binary_ev}, mass-loss rates from non-conservative mass transfer in our binary models exceed typical RSG mass-loss rates by several orders of magnitude. Hence, steady-state stellar wind mass-loss rates are expected to play a subdominant role in our models. For hotter stars, the Dutch scheme uses the recipe of \cite{Vink+2001AandA} or \cite{Nugis+2000AandA}, depending on whether the surface mass fraction of hydrogen is greater than 0.4 or not, respectively. 

Wind mass loss is enhanced due to rotation via multiplication of the factor $(1-\Omega/\Omega_{\mathrm{crit}})^{-0.43}$ \citep{Heger+2000ApJ}, where $\Omega$ and $\Omega_{\mathrm{crit}}$ are the equatorial rotational velocity and breakup velocity, respectively.

\subsubsection{Roche Lobe Overflow}  \label{subsubsec:rlof}

The donor evolves into a Red Supergiant (RSG) and develops an extended envelope in our binary models, which undergo Case B or Case C RLOF, depending on the evolutionary stage of the donor at the onset of RLOF \citep{Tauris+2006book}. The mass transfer rate during RLOF is calculated using the optically thick mass-transfer prescription by \cite{Kolb+1990AandA}, which is well suited for these systems, where the pressure scale height in the outer layers of the donor star is comparable to its radius.

Once RLOF is established, the subsequent accretion of mass and angular momentum onto the companion becomes a key ingredient of the binary evolution. This is a complex problem that is actively being explored with multidimensional hydrodynamic simulations \citep{MacLeod+2020ApJ_a,MacLeod+2020ApJ_b,Ryu+2025AandA}. In 1D stellar evolution calculations, this process must be described using simplified, physically motivated prescriptions. In this work, we adopt a disk-mediated accretion scenario \citep[e.g.,][]{Lu+2023MNRAS}. Material transferred through the inner Lagrange point is assumed to carry the specific angular momentum of a Keplerian orbit at the surface of the accretor \citep{deMink+2013ApJ}. This leads to rapid spin-up of the accretor toward critical rotation. We assume that mass transfer proceeds conservatively until the accretor reaches critical rotation, beyond which further accretion is inhibited, as additional angular momentum cannot be efficiently accommodated and the associated disk material becomes marginally unbound. The system then transitions to a non-conservative regime in which the accretor is maintained near critical rotation and excess transferred material is expelled from the system \citep{Packet1981AandA}. Accordingly, $\beta_{\rm eff}$, the effective mass transfer efficiency parameter \cite[as defined by][]{Tauris+2006book} is as follows:

\vspace{-3mm}
\begin{equation}   \label{eq:beta_eff}
\beta_{\mathrm{eff}} \equiv \dot{M}_{\mathrm{acc}}/\dot{M}_\mathrm{{tr}} = 
\begin{cases}
    1, & \Omega < \Omega_{\mathrm{crit}} \\
    0, & \Omega \geq \Omega_{\mathrm{crit}},
\end{cases}
\end{equation}

where $\dot{M}_{\mathrm{tr}}$ and $\dot{M}_{\mathrm{acc}}$ are the rates of mass transfer from the donor to the accretor and the rate of mass gained by the accretor, respectively. We model the mass lost from the system as an outflow through the outer Lagrange point (L2), adopting the velocity and specific angular momentum of the ejected material from the hydrodynamic simulations of \cite{Scherbak+2025ApJ}. They find that the outflow carries a specific angular momentum comparable to that at the L2 point. For mass ratios $q$ close to unity, as is the case for our relevant models (particularly close to core collapse of the donor), they report the specific angular momentum of the outflow to be $\approx 0.8h_{L2}$, where $h_{L2}$ is the specific angular momentum at the L2 point. The associated angular momentum loss is included self-consistently in the orbital evolution by removing mass with this specific angular momentum (set using the \texttt{mass\_transfer\_gamma} parameter in \mesa) once the accretor approaches critical rotation. This picture is broadly supported by observations of interacting binaries, which frequently exhibit evidence for non-conservative mass transfer. For example, purely conservative mass transfer cannot explain the current mass ratio and orbital period of all massive algols in the Milky Way and Magellanic Clouds \citep{Sen+2026ApJ}.

\subsubsection{Core collapse and SN}  \label{subsubsec:SN}

We track the evolution of both stars in the binary and their interaction as discussed above until the central temperature of the donor,  $\mathrm{log}_{10}\,T_c=9.1$\,K. For all of our models, this marks the end of the core carbon burning phase, with the core carbon fraction dropping below $10^{-6}$. Following this, we evolve the donor star to the point of core collapse with binary mass transfer turned off to overcome numerical instabilities. The interval between the end of carbon burning and core collapse is relatively short ($\sim$\,10 years for all our models), leaving insufficient time for either the donor's remaining lifetime or mass-loss rate to change significantly, even if binary mass transfer was included.

The explosion energy and the amount of $^{56}\mathrm{Ni}$ synthesized are set to $10^{51}$ ergs and $0.04M_{\odot}$, respectively, in accordance with the Zero Age Main Sequence (ZAMS) mass of our donor stars and the distributions reported in recent parametrized explosion studies \citep{Schneider+2021AandA,AD+2023AandA}. It has been shown that these explosion parameters are relatively insensitive to the progenitor's mass-loss history and prior binary evolution \citep{Pejcha+2015ApJ,Ertl+2016ApJ,Sukhbold+2016ApJ}.

\begin{deluxetable*}{cccccccccccc}
\tablecaption{Selected properties of the binary models.} 
\label{tab:mesa_models}
\tablehead{
\colhead{Model} & \colhead{$q_i$} & \colhead{$P_i$} & \colhead{RLOF} & \colhead{$a$} & \colhead{$M_f$} & \colhead{$R_f$} & \colhead{$M_{\mathrm{env,f}}$} & \colhead{$\dot{M}_{t,f}$} & \colhead{Mass lost} & \colhead{$X_s(H)$} & \colhead{Expected}\\[-2ex]
\colhead{name} &  & \colhead{(days)} & \colhead{Case} & \colhead{($R_{\odot}$)} & \colhead{($M_{\odot}$)} & \colhead{($R_{\odot}$)} & \colhead{($M_{\odot}$)} & \colhead{($M_{\odot}\mathrm{\,yr^{-1}}$)} & \colhead{($M_{\odot}$)} & & \colhead{SN type}\\[-2ex]
}
\decimalcolnumbers
\startdata
q95$\_$p26 & 0.95 & 2600 & C   & 3468 & 9.63 & 981  & 2.99      & $1.35\times10^{-4}$ & 7.06 & 0.609  & IIL\\
q75$\_$p26 & 0.75 & 2600 & C   & 3317 & 9.35 & 959  & 2.73      & $1.57\times10^{-4}$ & 6.60 & 0.608  & IIL\\
q55$\_$p26 & 0.55 & 2600 & C   & 2970 & 9.68 & 993  & 3.13      & $1.02\times10^{-3}$ & 6.26 & 0.608  & IIL\\
q95$\_$p22 & 0.95 & 2200 & B+C & 3322 & 7.88 & 753  & 1.27      & $3.74\times10^{-5}$ & 8.79 & 0.601  & IIb\\
q75$\_$p22 & 0.75 & 2200 & B   & 3158 & 7.07 & 507  & 0.47      & $0$                 & 8.88 & 0.573  & IIb\\
q95$\_$p08 & 0.95 & 800  & B   & 2548 & 6.09 & 1.30 & $10^{-7}$ & $0$                 & 10.68& 0.0002 & Ib\\
\enddata
\tablecomments{Columns are as follows: (1) Model name; also encodes initial accretor-to-donor mass ratio $q_i$ (col 2) and initial orbital period $P_i$ in days (col 3), (4) RLOF case, (5) final orbital separation, (6) final mass of donor, (7) final radius of donor, (8) final envelope mass of donor, (9) final RLOF mass transfer rate, (10) total mass lost from the binary system, (11) surface hydrogen abundance of donor, and (12) the SN type expected solely from the final mass of the hydrogen-rich donor envelope (not including spectroscopic CSM interaction sigantures). For all models, the ZAMS mass of the donor is $16M_{\odot}$.}
\end{deluxetable*}

\subsection{Hydrodynamic modeling of the CSM} \label{sec:csm}

Material lost from the binary forms CSM surrounding the system. Following the donor's subsequent SN explosion, the ejecta eventually interact with this CSM. We model the binary outflow and resulting CSM formation using the 3D expanding mesh hydrodynamics code, \sprout \citep{Mandal+2023_sprout}. We solve the equations of Newtonian hydrodynamics in Cartesian coordinates co-rotating with the binary, accounting for the binary gravitational potential (the Roche potential) and mass outflows. The equations are therefore as follows:

\begin{equation}
\label{eq:euler}
    \begin{gathered}
        \partial_t(\rho) + \nabla \cdot ( \rho \mathbf{v} ) = W_M \\
        \partial_t( \rho \mathbf{v} ) + \nabla \cdot ( \rho \mathbf{vv} + P \mathbf{I} ) = \rho \mathbf{a}_{\mathrm{R}} + \mathbf{W}_S \\
        \partial_t E + \nabla \cdot \left[ \left(E + P\right)\mathbf{v} \right] = \rho \mathbf{a}_{\mathrm{R}}\cdot\mathbf{v} + W_E,
    \end{gathered}
\end{equation}

where $\rho$, $P$, $\mathbf{v}$, and $E\equiv \rho v^2/2 + \rho e$ are the density, pressure, velocity, and total (kinetic+internal) energy density of the fluid (in the co-rotating frame),  respectively. $\mathbf{I}$ is the identity tensor, and $\mathbf{a}_{\mathrm{R}}$ is the acceleration due to the Roche potential. $W_M$, $\mathbf{W}_S$ and $W_E$ are the mass, momentum and energy source terms for the outflows (stellar winds and the L2 outflow). The acceleration term $\mathbf{a}_{\mathrm{R}}$ supplies the momentum and energy contributions due to the binary's gravity and the rotating frame (the centrifugal and Coriolis acceleration), expressed as follows:

\begin{equation}  \label{eq:Roche_accel}
    \mathbf{a}_{\mathrm{R}} 
    = -\frac{GM_1}{\left|\mathbf{r_1} \right|^3}\mathbf{r_1}
    -\frac{GM_2}{\left|\mathbf{r_2} \right|^3}\mathbf{r_2}
    \\ -\mathbf{\Omega}\times(\mathbf{\Omega} \times \mathbf{r})
    + 2(\mathbf{\Omega} \times \mathbf{v}),
    %\Phi(\mathbf{r}) = -\frac{GM_1}{\left| \mathbf{r} - \mathbf{r_1} \right|} - \frac{GM_2}{\left| \mathbf{r} - \mathbf{r_2} \right|} - \frac{1}{2}\left| \left(\mathbf{r}-\mathbf{r_{CM}}\right)\times\mathbf{\Omega} \right|^2,
\end{equation}

where $\mathbf{r_1}$, $\mathbf{r_2}$ and $\mathbf{r}$ are the position vectors with respect to the donor, the accretor, and the center of mass, respectively. $M_1$ and $M_2$ are the masses of the donor and the accretor, respectively. $\mathbf{\Omega}=(0,0,\Omega)$ is the angular velocity, thus defining the x-y plane to be the binary orbital plane. $\Omega=\left[G(M_1+M_2)/a^3\right]^{1/2}$ is the angular speed of the binary, with $a$ being the binary separation. The acceleration $\mathbf{a}_{\mathrm{R}}$ is calculated as the negative of the gradient of the Roche potential $\Phi_R$, which is as follows:

\begin{equation}  \label{eq:Roche_pot}
    \Phi_R
    = -\frac{GM_1}{\left|\mathbf{r_1} \right|}  -\frac{GM_2}{\left|\mathbf{r_2} \right|} - \frac{1}{2}\Omega^2(\mathbf{r}-\mathbf{r}\cdot\hat{k})^2.
\end{equation}

We limit the minimum value of $\Phi_R$ to be $\Phi_{min} = -\mathrm{G\cdot max}(M_1,M_2)/\delta$, where $\delta$ is a small distance compared to the computational domain. This effectively sets the acceleration term $\mathbf{a}_{\mathrm{R}}$ to zero in the vicinity of either star and avoids the issue of infinite acceleration near point-like gravity sources.

The source terms corresponding to mass loss from the binary system must take into account stellar winds from the surface of the donor and the accretor, as well as a marginally unbound outflow from the L2 point, as discussed in Section~\ref{subsec:mesa}. Wind mass-loss rates from the donor and the accretor are calculated by \mesa. The mass-loss rate from L2 follows from \autoref{eq:beta_eff}:

\vspace{-2mm}
\begin{equation}
\label{eq:M_csm}
        \dot{M}_{\mathrm{L2}} = (1-\beta_{\mathrm{eff}})\dot{M}_\mathrm{{tr}}.
\end{equation}

This mass is injected at the L2 point with radially outward velocity equal to one percent of the orbital velocity at L2 ($v_{r}=0.01r_{\rm L2}\Omega$), where $r_{\rm L2}$ is the distance between the L2 point and the center of mass, computed using the Roche lobe calculator by \cite{Leahy+2015ComAC}. Following \cite{Scherbak+2025ApJ}, we assign the material injected at L2 a specific angular momentum $h_{\mathrm{loss}}=0.8h_{\mathrm{L2}}=0.8r_{\rm L2}^2\Omega$ (in the inertial frame), appropriate for the near-unity mass ratios found for our models at late evolutionary stages. The winds from the donor and the accretor are injected as radially symmetric outflows in the co-rotating frame. The initial wind speed is proportional to the escape velocity at the star's surface, with an order of unity proportionality constant \citep{Lamers+1995ApJ}. Since the mass-loss rates are less than the L2 mass-loss rate by several orders of magnitude, the stellar winds are found to have a minimal impact on our hydrodynamic models.

The equations are closed with a polytropic equation of state: $P = \left(\gamma-1\right)\rho e$, taking $\gamma=5/3$. The computational domain is initiated with a size similar to that of the binary separation, $a$, but is gradually expanded so that each side of the domain reaches a final length of $10^{16}$ cm. This size corresponds to the maximum extent of the CSM included in the \stella models.

\subsection{Computation of SN light curves} \label{sec:stella}

Following core collapse, the donor star models are evolved using \mesa until the SN shock wave reaches the vicinity of the stellar surface, producing pre-SN ejecta profiles \citep[as in][]{Paxton2018}. For each binary model, we also sample the 3D CSM (modeled using \texttt{Sprout}) along three viewing angles, $\theta=0^{\circ},45^{\circ}\mathrm{\,and\,}90^{\circ}$, with $\theta$ being the angle between the binary axis and the line of sight. This yields a set of angle-dependent 1D CSM profiles.

Each 1D CSM profile is appended to the outer boundary of the ejecta profile and used as input for \stella, which solves the time-dependent equations of radiation hydrodynamics in the multi-group approximation with a two-temperature (gas and radiation) treatment. The \stella models track the interaction between the ejecta and the CSM and compute the emergent multi-band optical light curves. Each model is evolved for 120 days after explosion.

\section{Results} \label{sec:results}

In this section, we describe the evolution of our binary models, along with the structure of the CSM. Thereafter, we present light curves of the ensuing CCSNe, and discuss how they vary with viewing angle.

\subsection{Binary evolution}
\label{subsec:binary_ev}

The observed properties of CCSNe depend on both the pre-SN structure of the progenitor and its immediate circumstellar environment. In this study, the explosion energy and $^{56}$Ni mass are held fixed, while we investigate the impact of key parameters such as the final envelope mass, $M_{\mathrm{env,f}}$, and the mass-loss rate from the system in the final few centuries prior to core collapse. Together, these variables set the diffusion timescale and hydrogen content of the ejecta, as well as the degree of ejecta--CSM interaction. In binary systems, both of these quantities are primarily determined by the initial orbital period, $P_i$, and mass ratio,  $q_i$. In systems with shorter periods ($P_i\lesssim2200$ days), RLOF begins after core hydrogen depletion but before helium ignition \citep[i.e., Case B RLOF;][]{Tauris+2006book}. Such systems lose most of their hydrogen-rich envelope, whereas wider systems (the q**\_p26 models, with $P_i = 2600$ days\footnote{Henceforth, we use the notation q**\_p26 to denote all \\ models with a period of 2600 days, and so on.}) undergo Case C RLOF (post core helium depletion) and retain more massive envelopes. In some Case B systems, however, RLOF is initiated earlier, resulting in the retention of some portion of the donor envelope ($M_{\mathrm{env}}\gtrsim0.3M_\odot$). The presence of a substantial envelope leads to a later RLOF phase (Case C), as shown by \cite{Ercolino+2024AandA}. This behavior is found for our model q95\_p22. Table~\ref{tab:mesa_models} shows that the models exhibit a monotonic decrease in $M_{\mathrm{env,f}}$ with decreasing $P_i$, as demonstrated previously \citep{Ercolino+2024AandA,Matsuoka+2024ApJ}.

In terms of the eventual explosion of the donor, this trend maps onto a continuum of CCSN types, governed by the residual hydrogen-rich envelope mass alone. Donors with massive envelopes produce Type IIP SNe, where hydrogen recombination sustains a plateau in the light curve. Partially stripped envelopes lead to Type IIL SNe with faster post-peak decline, and strongly stripped systems with low $M_{\mathrm{env,f}}$ produce SESNe (Type IIb or Ib), where the paucity of hydrogen results in weak or absent hydrogen lines \citep[][also see Section~\ref{subsec:light_curves_no_csm}]{Dessart+2024AandA}. Note that these SN subtypes are expected solely on the basis of the residual hydrogen-rich envelope mass, and can change in the presence of CSM around the progenitor \citep{Smith2017hsn..book}.

\begin{figure}
\vspace{2mm}
\centering
\includegraphics[width=0.48\textwidth]{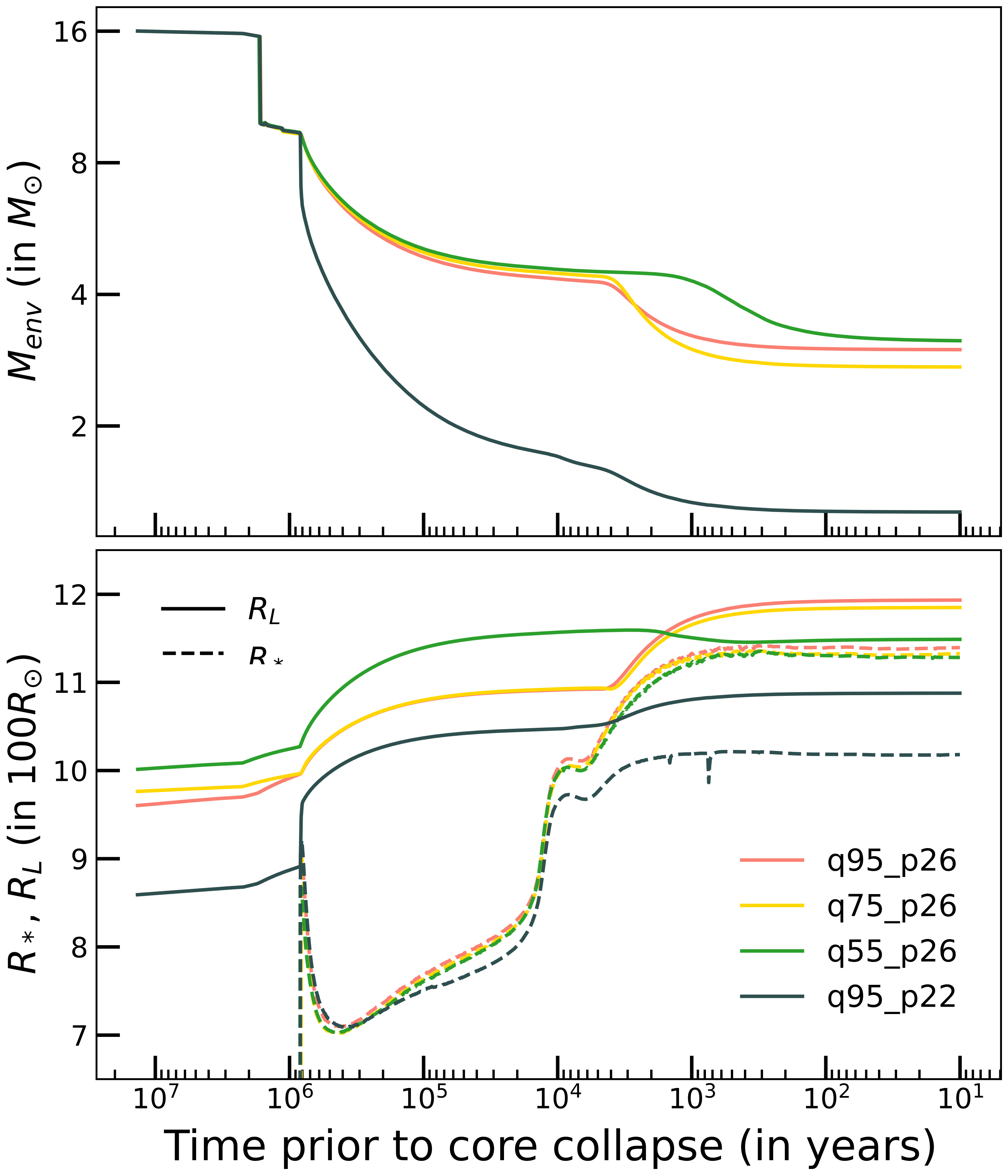}
\caption{Evolution of donor star parameters in binary systems undergoing Case C (post helium burning) Roche lobe overflow (RLOF). \textit{Top}: Envelope mass of the donor as a function of time prior to core collapse. \textit{Bottom}: Time evolution of the donor stellar radius ($R_*$, dashed lines) and donor's Roche lobe radius ($R_L$, solid lines). RLOF begins when the donor radius approaches the Roche lobe radius.}
\label{fig:M_R_ev}
\end{figure}

The dependence of $M_{\mathrm{env,f}}$ on $q_i$, on the other hand, is non-monotonic (see Table~\ref{tab:mesa_models}). This behavior is illustrated in \autoref{fig:M_R_ev}, which depicts the time evolution of $M_{\mathrm{env}}$ (top panel), along with the donor star radius, $R_*$, and Roche lobe radius, $R_{\rm RL}$, (bottom panel) for all systems undergoing Case C or Case B + Case C RLOF. Lower mass ratios lead to larger Roche lobe radii \citep{Eggleton1983ApJ}, delaying the onset of RLOF (\autoref{fig:M_R_ev}), and thereby reducing the time available for envelope stripping. At the same time, the response of the Roche lobe during mass transfer depends sensitively on the initial mass ratio, $q_i$. While $R_{\rm RL}$ increases following contact for higher $q_i$ ($=0.75$ or $0.95$), it decreases for sufficiently low $q_i$ ($0.55$ in this case). For a more detailed study of the dependence of the donor's envelope mass on binary parameters (including Case B and Case C RLOF systems), we refer the reader to \cite{Ercolino+2024AandA}.

\begin{figure}
\centering
\includegraphics[width=0.48\textwidth]{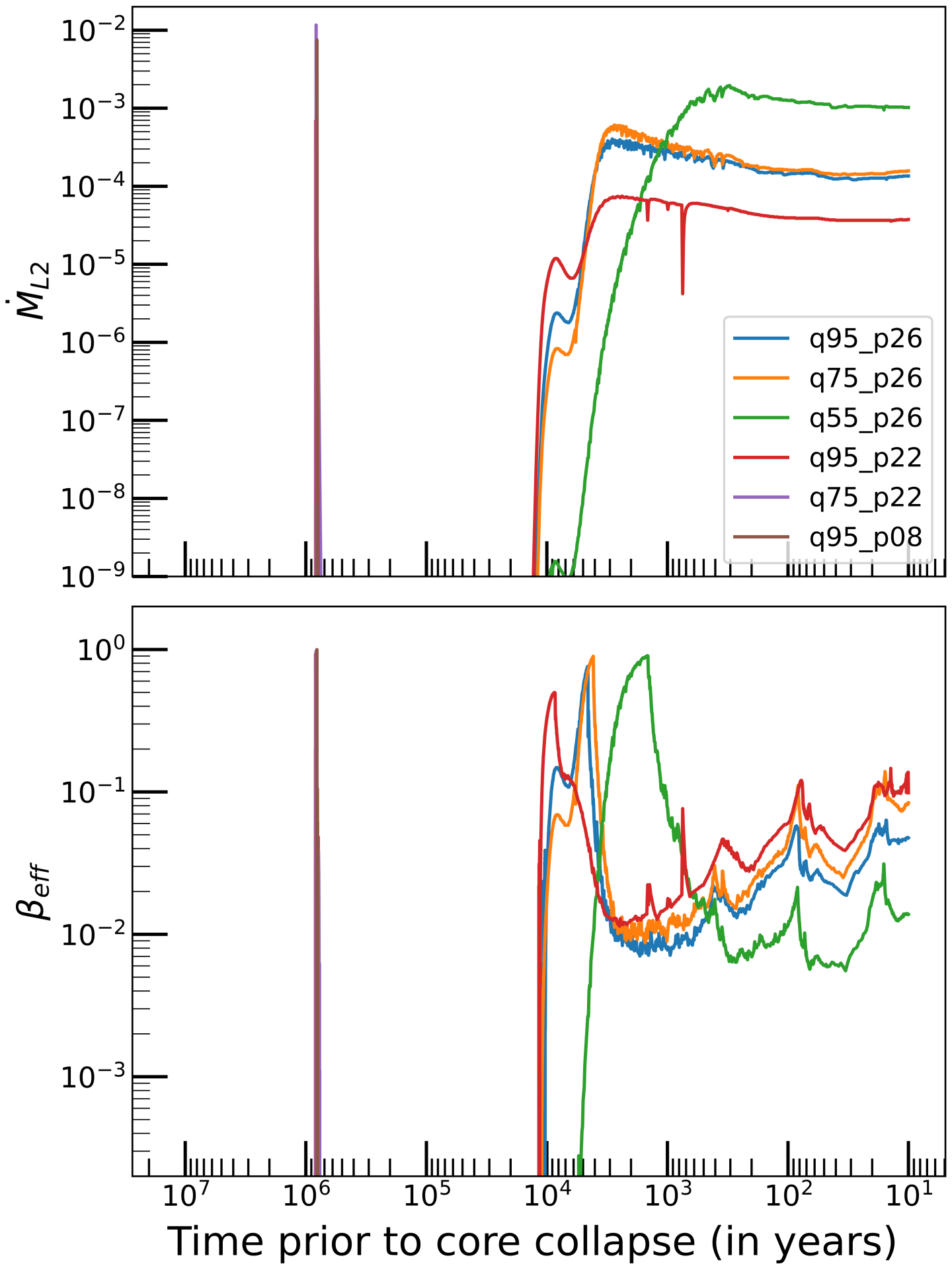}
\caption{\textit{Top}: Mass-loss rate from the outer Lagrange point L2 ($\dot{M}_{\mathrm{L2}}$) in our models, as a function of time prior to core collapse of the donor. The plotted mass-loss rate corresponds to the material lost into the circumbinary environment and used to construct the CSM in the hydrodynamic models. Systems undergoing Case B mass transfer (q75\_p22, q95\_p22, and q95\_p08) lose mass earlier ($\sim10^6$ years before core collapse of the donor), whereas, Case C systems exhibit enhanced mass loss closer to core collapse, resulting in dense CSM close to the progenitor. \textit{Bottom}: The effective mass-transfer efficiency ($\beta_{\mathrm{eff}}$) as a function of time prior to core collapse of the donor, where $\beta_{\mathrm{eff}}=1$ indicates all mass lost from the donor via RLOF is accreted onto the companion, and $\beta_{\mathrm{eff}}=0$ implies that all lost mass leaves the binary system through the L2 point.}
\label{fig:mdot_ev}
\end{figure}

The shrinking Roche lobe relative to the stellar radius enhances the mass-transfer rate \citep[e.g., Eq.~13 of][]{Paxton2015}, leading to a complex, non-monotonic dependence of $M_{\mathrm{env,f}}$ on $q_i$. This behavior can be understood in terms of the efficiency of mass accretion by the companion. For low mass ratios, the accretor is both less massive and more easily spun up to critical rotation, limiting its ability to accept transferred material. As a result, mass transfer becomes highly non-conservative, and a large fraction of the overflowing gas is expelled from the system. This is quantified by the effective mass-transfer efficiency $\beta_{\mathrm{eff}}$, shown in the top panel of \autoref{fig:mdot_ev}. In all models, $\beta_{\mathrm{eff}}$ briefly approaches unity at the onset of RLOF, but rapidly declines to $\lesssim10\%$ as the accretor reaches critical rotation \citep{Packet1981AandA}. The effect is most pronounced in the q55\_p26 system, where $\beta_{\mathrm{eff}} \approx 1\%$, indicating that nearly all transferred mass is lost from the system. The associated loss of angular momentum causes the Roche lobe radius to shrink following contact. This, in turn, enhances the mass-transfer rate, establishing a positive feedback loop that is unique to the lowest-$q_i$ system in our sample. For even lower mass ratios, this feedback is expected to drive unstable mass transfer on a dynamical timescale \citep[e.g. CEE;][]{Ercolino+2024AandA}. An in-depth investigation of this effect is beyond the scope of the present work.

The mass-loss rate from the system defines the CSM structure. In our models, the dominant contribution comes from outflows through the L2 point, which exceeds wind mass-loss rates from either the donor or the accretor by several orders of magnitude. In \autoref{fig:mdot_ev} (top panel), we plot the evolution of the L2 mass-loss rate ($\dot{M}_{\mathrm{L2}}$; see \autoref{eq:M_csm}) for all models. These rates are consistent with previous findings that Case C RLOF can drive sustained, late-stage mass loss up to the onset of core collapse \citep{Matsuoka+2024ApJ,Ercolino+2024AandA}. In contrast, systems undergoing Case B RLOF (i.e., the shorter period models q95\_p08, q75\_p22, and q95\_p22) lose mass much earlier. This occurs typically $10^6$ years prior to core collapse of the donor, during hydrogen shell burning (seen as short duration spikes for corresponding plots in \autoref{fig:mdot_ev}. It is worth highlighting that Case B RLOF is more effective at stripping the donor and causes more mass loss from the system, as seen in column 10 of Table~\ref{tab:mesa_models}. The bottom panel of \autoref{fig:mdot_ev} shows that the companion becomes largely inefficient at accreting mass from the donor once RLOF is initiated, with the effect strongest for the model with the least massive companion (q55\_p26).

The dichotomy in the mass-loss history of binary CCSN progenitors translates directly into differences in the spatial distribution of the CSM, and hence the timing of the onset of ejecta-CSM interaction. Immediate CSM interaction (such as in Type IIn SNe) would be expected in systems that undergo stable Case C RLOF \citep{Tsai+2026arXiv} or experience unstable mass transfer on dynamical timescales (such as a CEE event) near the end of the progenitor’s life. By contrast, mass loss solely due to Case B RLOF is expected to produce CSM located at large radii by the time of explosion. As a result, interaction with this CSM, if any, may occur only at much later times ($\gtrsim10^3$ years). Thus, the binary interaction history plays a central role in determining the pre-core collapse envelope mass and the mass-loss rate, shaping the diversity of CCSN light curves and spectral types.

\begin{figure*}
\centering
\includegraphics[width=0.98\textwidth]{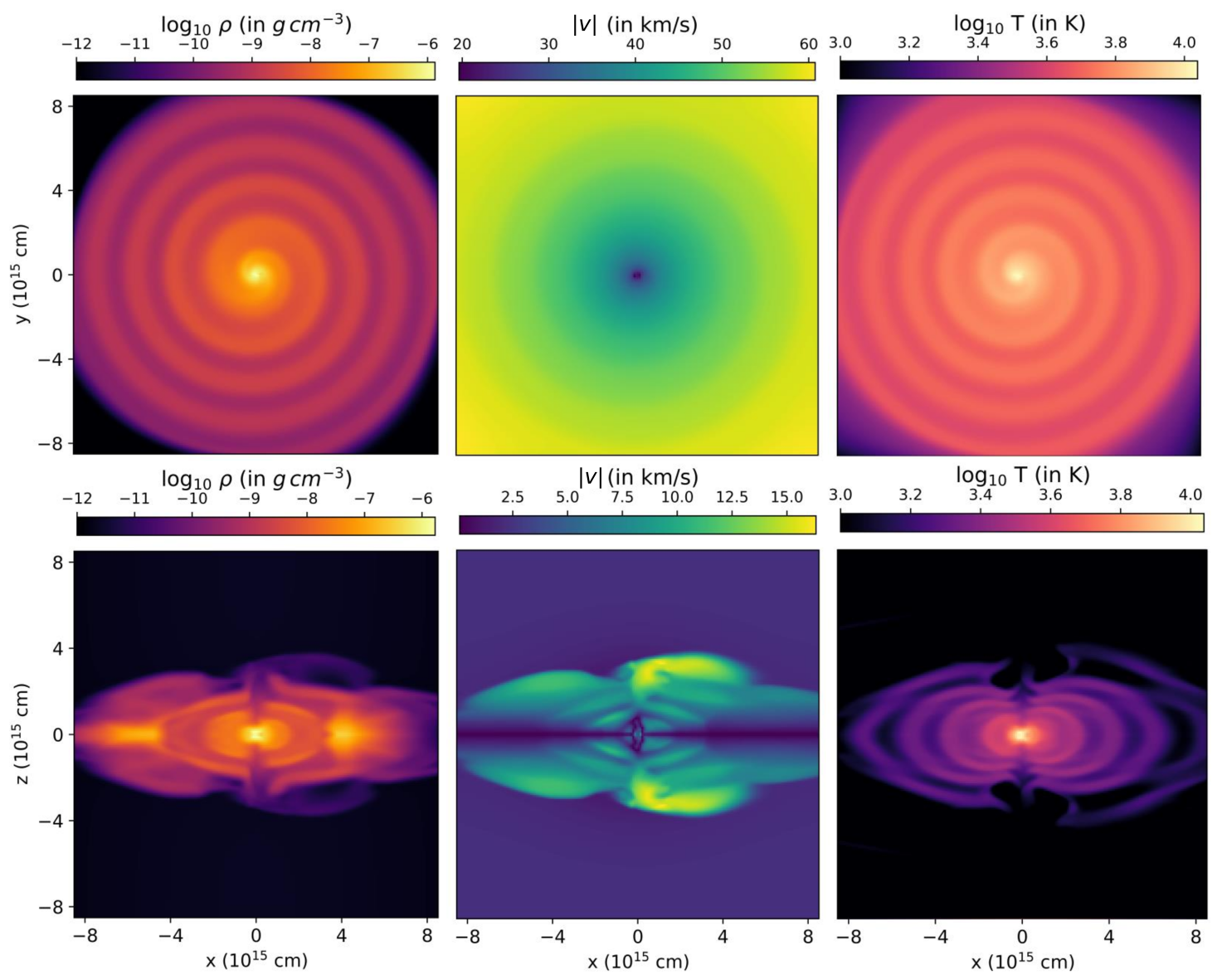}
\vspace{-0mm}
\caption{2D density (\textit{left panels}), velocity magnitude (\textit{middle panels}), and temperature (\textit{right panels}) slices for q55\_p26 hydrodynamic CSM model, shown for both the equatorial plane (x-y plane, \textit{top panels}) and the meridional plane (x-z plane, \textit{bottom panels}). These slices were extracted after the model reached a quasi-steady state (see Section~\ref{subsec:csm_structure}). The angular momentum and gravitational potential of the binary system generates an equatorially-enhanced circumstellar structure from the L2 outflow, with spiral patterns appearing in the equatorial plane \citep{Shu+1979ApJ}.}
\label{fig:csm_slice}
\end{figure*}

\subsection{Structure of the CSM}
\label{subsec:csm_structure}

%\begin{figure} \centering \includegraphics[width=0.48\textwidth]{csm_eqt_5526.png} \includegraphics[width=0.48\textwidth]{csm_mer_5526.png} \caption{Density slices in the equatorial plane (x-y plane, \textit{top panel}) and the meridional plane (x-z plane, \textit{bottom panel}) from a hydrodynamic model corresponding to the q55\_p26 system. The angular momentum and gravitational potential of the binary system generates an equatorially-enhanced circumstellar structure from the L2 outflow, with spiral patterns appearing in the equatorial plane \citep{Shu+1979ApJ}.}\label{fig:csm_slice} \end{figure}

The hydrodynamic binary outflow models are run for ten orbits. By this time, sufficient mass transfer has occurred to establish a quasi-steady state in our computational domain. The resulting structures are shown in \autoref{fig:csm_slice}, where we plot midplane (x-y and x-z) slices of the CSM corresponding to the density, velocity and temperature in the q55\_p26 model. Material lost through the outer Lagrange point forms a strongly aspherical outflow concentrated toward the orbital plane. The outflow follows a prominent spiral trajectory \citep{Shu+1979ApJ}, similar to numerical models of L2 mass loss from binary systems \citep{Pejcha+2016MNRAS,Scherbak+2025ApJ}. The meridional slices reveal the formation of a geometrically extended, equatorially enhanced structure with comparatively low densities in the polar regions, with density contrasts between equatorial and polar directions reaching two orders of magnitude at a distance of $3\times10^{15}$ cm (approximately ten times the orbital separation) from the binary.

\begin{figure*}
\centering
\includegraphics[width=0.98\textwidth]{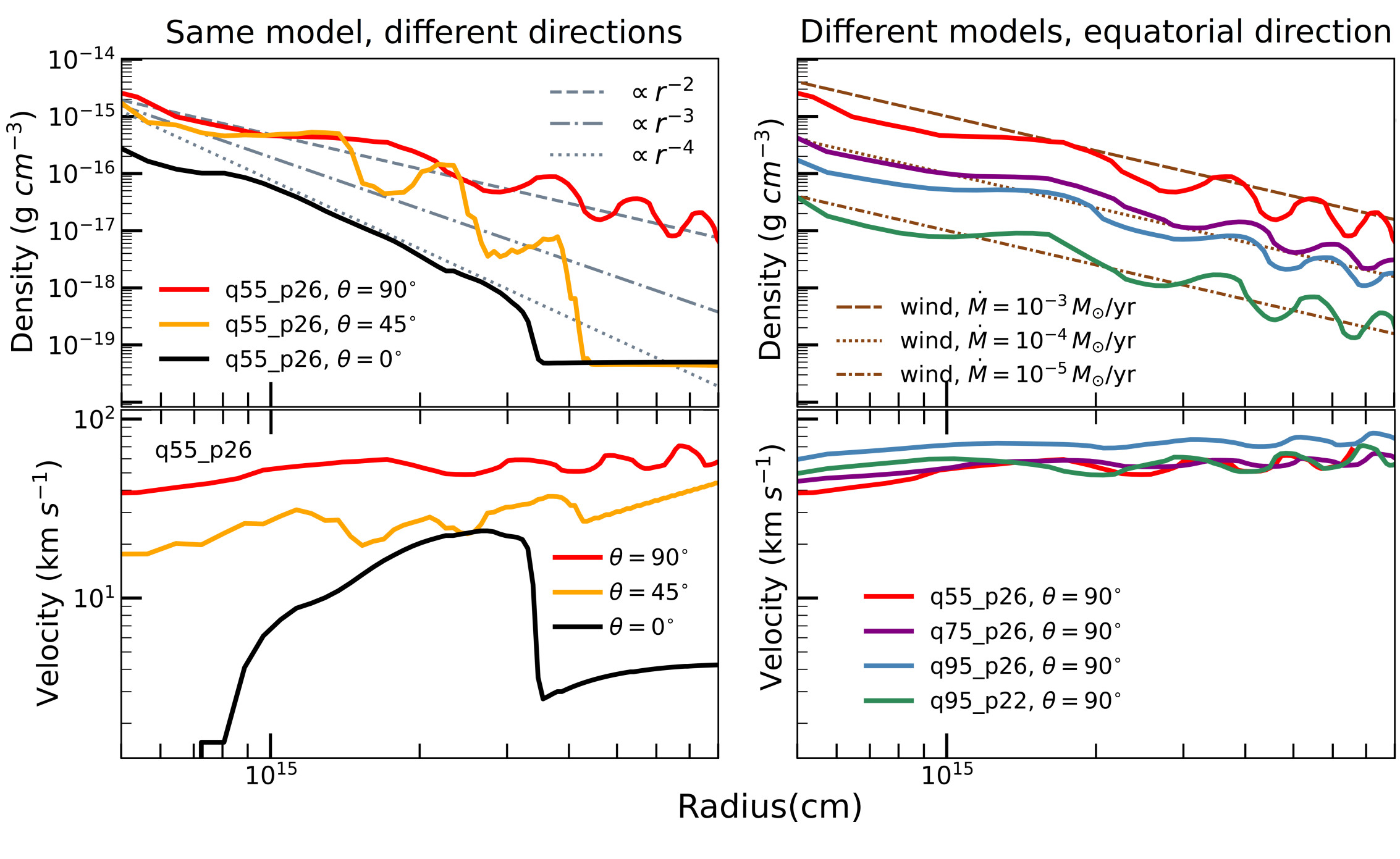}
\vspace{-5mm}
\caption{Direction-dependent 1D hydrodynamic profiles of our CSM models. \textit{Top left}: density profiles for the q55\_p26 models in the polar, oblique, and equatorial directions ($\theta=0^{\circ},45^{\circ},\mathrm{and\,}90^{\circ}$ respectively, where $\theta$ is the angle between the observer's line of sight and the binary axis). The density profiles become steeper overall (particularly at large distances) as one moves away from the equatorial plane towards the binary axis. \textit{Top right}: equatorial density profiles for all our hydrodynamic models. These are found to be overall consistent with a steady wind-like density profile, with the mass loss rate being that from the L2 outflow (see Table~\ref{tab:mesa_models}) and the wind speed being $50\mathrm{\,km\,s^{-1}}$. \textit{Bottom left}: velocity profiles for the q55\_p26 models along the three observer directions. \textit{Bottom right}: equatorial velocity profiles for all models. Change of viewing direction is seen to cause relatively larger variation in the CSM velocity compared to varying binary model parameters.}
\label{fig:csm_1D_slice}
\end{figure*}

%\gridline{\fig{csm_den_slices_5526.png}{0.49\textwidth}{}\fig{csm_den_slices_all.png}{0.49\textwidth}{}}
%\gridline{\fig{csm_vel_slices_5526.png}{0.49\textwidth}{}\fig{csm_vel_slices_all.png}{0.49\textwidth}{}}

The density and velocity contrast between different directions is further elucidated in \autoref{fig:csm_1D_slice}, where we plot 1D azimuthally averaged profiles extracted from the hydrodynamic CSM models. The q55\_p26 profiles (top left panel) demonstrate that the CSM density varies with the observing direction. The profile in the equatorial direction follows that for a steady wind ($\rho\propto r^{-2}$), as found commonly for L2 outflows \citep{Pejcha+2016MNRAS,Scherbak+2025ApJ}. Steeper density profiles are encountered in directions away from the orbital plane. For example, the density profile in the polar direction ($\theta=0^{\circ}$) for the q55\_p26 model is consistent with a $r^{-2}$ profile for relatively small radii ($r\lesssim8\times10^{14}$ cm), but gradually steepens to a $r^{-4}$ profile at larger radii. Similar behavior is found for all our models with substantial L2 mass loss. Such steep density profiles are, in fact, often inferred for interacting SNe \citep{Chandra+2015ApJ,Maeda+2022ApJ,Nagao+2023AandA,BW+2025ApJ_b} and attributed to eruptive mass loss, dynamical timescale events such as stellar mergers \citep{Tsuna+2024OJAp}, or accelerated winds \citep{Moriya+2018MNRAS,Aryan+2025ApJ}. Our results show that geometry can also play an important role in producing such apparently steep density profiles, even from steady outflows. In addition, as the top left panel of \autoref{fig:csm_1D_slice} shows, the extent or `outer radius' of the CSM is very large in the equatorial direction, making such viewing angles ideal for observing persistent signatures of ejecta-CSM interaction. In contrast, lines of sight not aligned with the orbital plane (say $\theta=0^{\circ}$ or $45^{\circ}$) exhibit a smaller extent of CSM (about a few times $10^{15}$ cm). One would thus expect relatively short-lived signs of interaction in such cases.

As expected from Table~\ref{tab:mesa_models}, we also recover varying mass-loss rates (spanning two orders of magnitude) from our CSM models, as shown by the equatorial density profiles from all our models with significant pre-explosion mass loss (\autoref{fig:csm_1D_slice}, top right panel). The spiral outflows produce undulations in the density profile, with the density contrast of a few between the crests and troughs of the undulations. As we will see in section~\ref{subsec:light_curves_with_csm}, these undulations can produce bumpy light curves reminiscent of a small subclass of Type IIn SNe \citep{Smith2017hsn..book}.

%\add{Another feature of interest is the undulation in the density profile along the equatorial line of sight, also apparent for all other CSM models (top right panel). NOTE FOR SHAZRENE: why are they important? \cite{Maercker2012}}

Strong contrasts are also found amongst the velocity profiles in different directions, as would be expected due to the high specific orbital momentum carried away by the outflow. This is seen from the  q55\_p26 velocity profiles  along different lines of sight (lower left panel of \autoref{fig:csm_1D_slice}). Velocities in the polar direction are found to reach less than $10\,\mathrm{km\,s^{-1}}$ compared to $40-60\,\mathrm{km\,s^{-1}}$ velocities in the equatorial direction. The differences in velocities among different models (lower right panel) is relatively less, expected from the relatively low variation in the final orbital period (and hence the specific angular momentum) in our models. Our results thus suggest geometric effects can alter the CSM speed by a factor of a few tens, resulting in large uncertainties in the CSM densities and mass-loss rates inferred for Type II SNe. 

%This implies that the ``true'' CSM speed for an object may be quite uncertain, and in fact may be difficult to accurately diagnose in many cases.

\subsection{Light curves assuming no CSM}
\label{subsec:light_curves_no_csm}

Prior to examining the effects of binary-generated CSM, we first establish a reference set of light curves by exploding the \mesa progenitor models directly in \stella, without appending the \sprout CSM profiles. These models provide both a baseline for later comparison and a validation that our progenitors produce the expected CCSN subtypes in the absence of ejecta-CSM interaction. In the absence of CSM, the observable SN subtype is largely determined by the hydrogen envelope mass at core collapse. Progenitors retaining massive hydrogen envelopes ($\sim7\,M_{\odot}$) produce Type IIP-like events with extended plateaus, while progressively lower envelope masses lead to Type IIL ($\sim0.3\,M_{\odot}$ to a few $M_{\odot}$), IIb (envelope mass in the range $\sim0.001-0.3\,M_{\odot}$)\footnote{This mapping between hydrogen envelope mass and CCSN subtype is approximate and depends on the adopted models.}, and ultimately Ib SNe as the hydrogen envelope is reduced or removed entirely \citep{Morozova+2015ApJ,Eldridge+2018PASA,Dessart+2019AandA,Hillier+2019AandA,Hiramatsu+2021ApJ,Dessart+2024AandA}. Based on the envelope masses listed in Table~\ref{tab:mesa_models}, our models are therefore expected to span a continuum of partially to strongly stripped CCSNe.

\begin{figure*}
\centering
\includegraphics[width=0.98\textwidth]{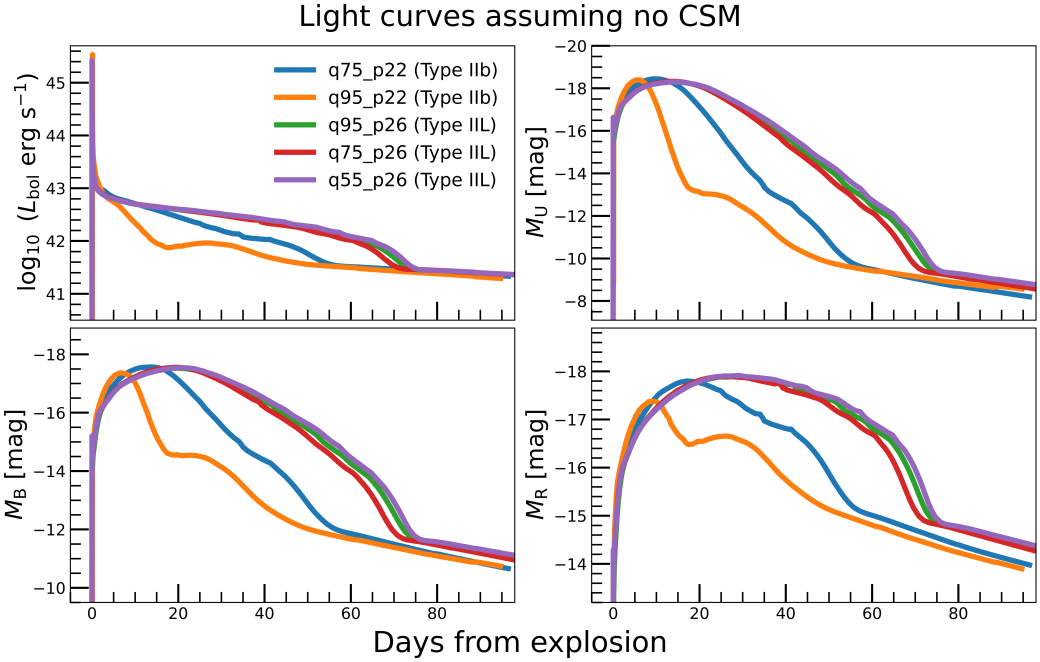}
\caption{Bolometric (\textit{top left}), $U$-band (\textit{top right}), $B$-band (\textit{bottom left}), and $R$-band (\textit{bottom right}) light curves from our models in the absence of CSM. The light curves span a continuum of CCSNe subtypes (Types IIL, IIb and Ib) depending on the envelope mass of the progenitor at core collapse (see Table~\ref{tab:mesa_models}).}
\label{fig:lc_no_csm}
\end{figure*}

The bolometric, $U$, $B$, $V$, and $R$-band light curves from our models exploded without CSM are presented in \autoref{fig:lc_no_csm}. The light curves span a range of decline rates and are qualitatively consistent with Types IIL and IIb SNe expected from their final envelope masses. Similar trends were reported by \cite{Eldridge+2018PASA} and \cite{Dessart+2024AandA}. Consistent with their study, we find that decreasing the initial orbital period leads to progressively stronger envelope stripping and therefore different SN subtypes. For example, models q**\_p26 produce Type IIL events, whereas models q**\_p22 produce Type IIb events, owing to lower orbital periods for the latter group. Light curves corresponding to the most stripped progenitor (model q95\_p08) could not be obtained due to numerical issues associated with photon leakage from an ultra-stripped stellar envelope.

This sequence is primarily controlled by the hydrogen envelope mass at core collapse (Table~\ref{tab:mesa_models}), which decreases significantly with orbital period. By contrast, varying the mass ratio at fixed orbital period produces comparatively minor differences in the light curves within our model grid. This suggests that, within our model grid, the initial orbital period is the primary parameter controlling envelope stripping and the resulting SN subtype (in the absence of CSM), while the mass ratio plays a comparatively secondary role. Having established the intrinsic SN emission in the absence of ejecta-CSM interaction, we now investigate how anisotropic binary-generated CSM modifies the observed light curves.

\subsection{Photosphere evolution in the presence of CSM}
\label{subsec:photosphere_ev_csm}

\begin{figure}
\centering
\includegraphics[width=0.48\textwidth]{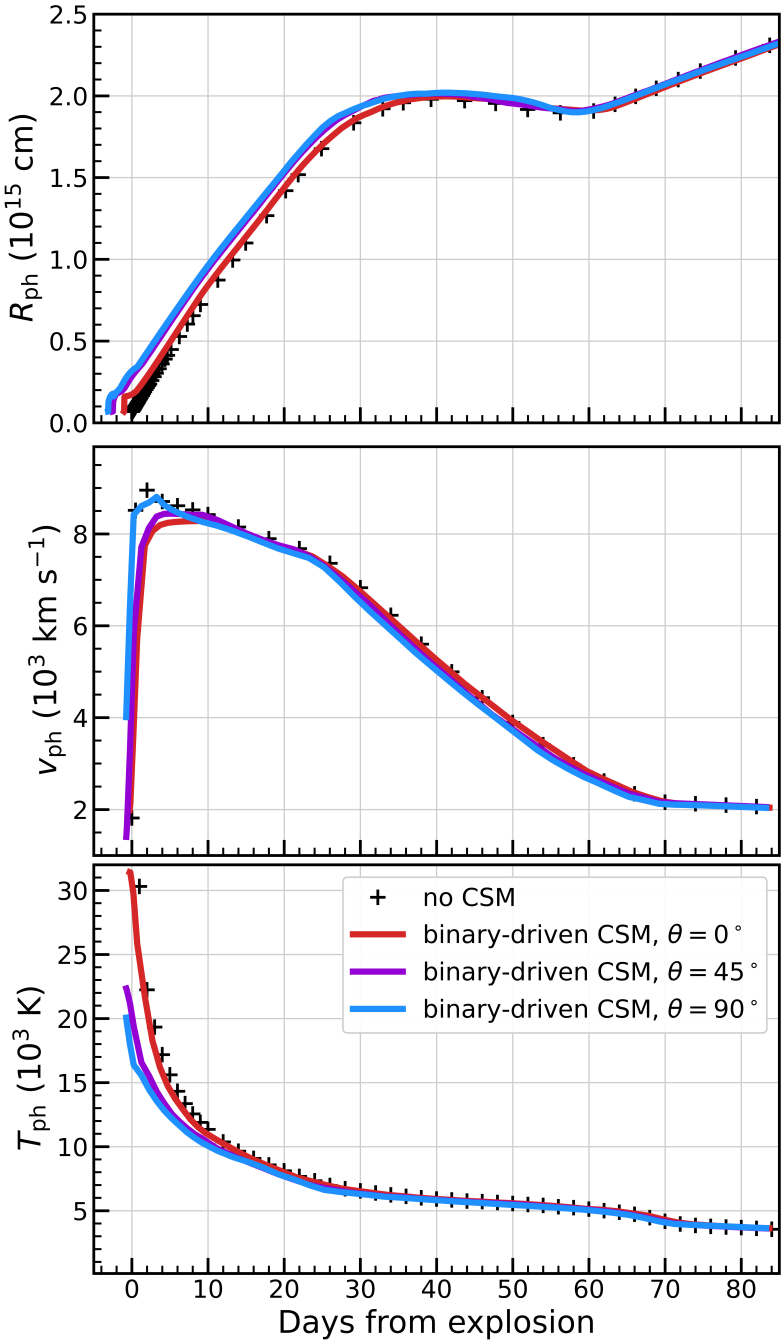}
\caption{Evolution of photosphere radius (\textit{top panel}), velocity (\textit{middle panel}) and temperature (\textit{bottom panel}) in the q55\_p26 SN explosion models with binary-driven CSM as in \autoref{fig:csm_slice}. Photospheric properties corresponding to all three observer directions ($\theta=0^{\circ},45^{\circ},\mathrm{and\,}90^{\circ}$) converge by the epoch of recombination ($T_{\rm ph}=6000$ K, $t\approx30$ days). Photospheric properties at early phases imply that the photosphere resides in the expanding ejecta instead of the CSM, even in directions with the highest CSM column density (see section~\ref{subsec:photosphere_ev_csm}).}
\label{fig:photosphere}
\end{figure}

The emergence of radiation from SN ejecta interacting with an equatorially-enhanced CSM is a fundamentally multi-dimensional phenomenon. It is thus useful to assess the validity of our direction-dependent 1D radiation hydrodynamic models before deriving observables. We do this by examining the evolution of the photospheric radius $R\mathrm{_{ph}}$, velocity $v\mathrm{_{ph}}$, and temperature $T\mathrm{_{ph}}$,  along different lines of sight, as plotted in \autoref{fig:photosphere} for the representative q55\_p26 model. Amongst our sample, this model achieves the highest pre-core collapse mass-loss rate  ($\approx10^{-3}\,M_{\odot}\,\mathrm{yr}^{-1}$), and is therefore expected to exhibit the largest variation in optical depth along different lines of sight. 

\begin{figure*}
\centering
\includegraphics[width=0.98\textwidth]{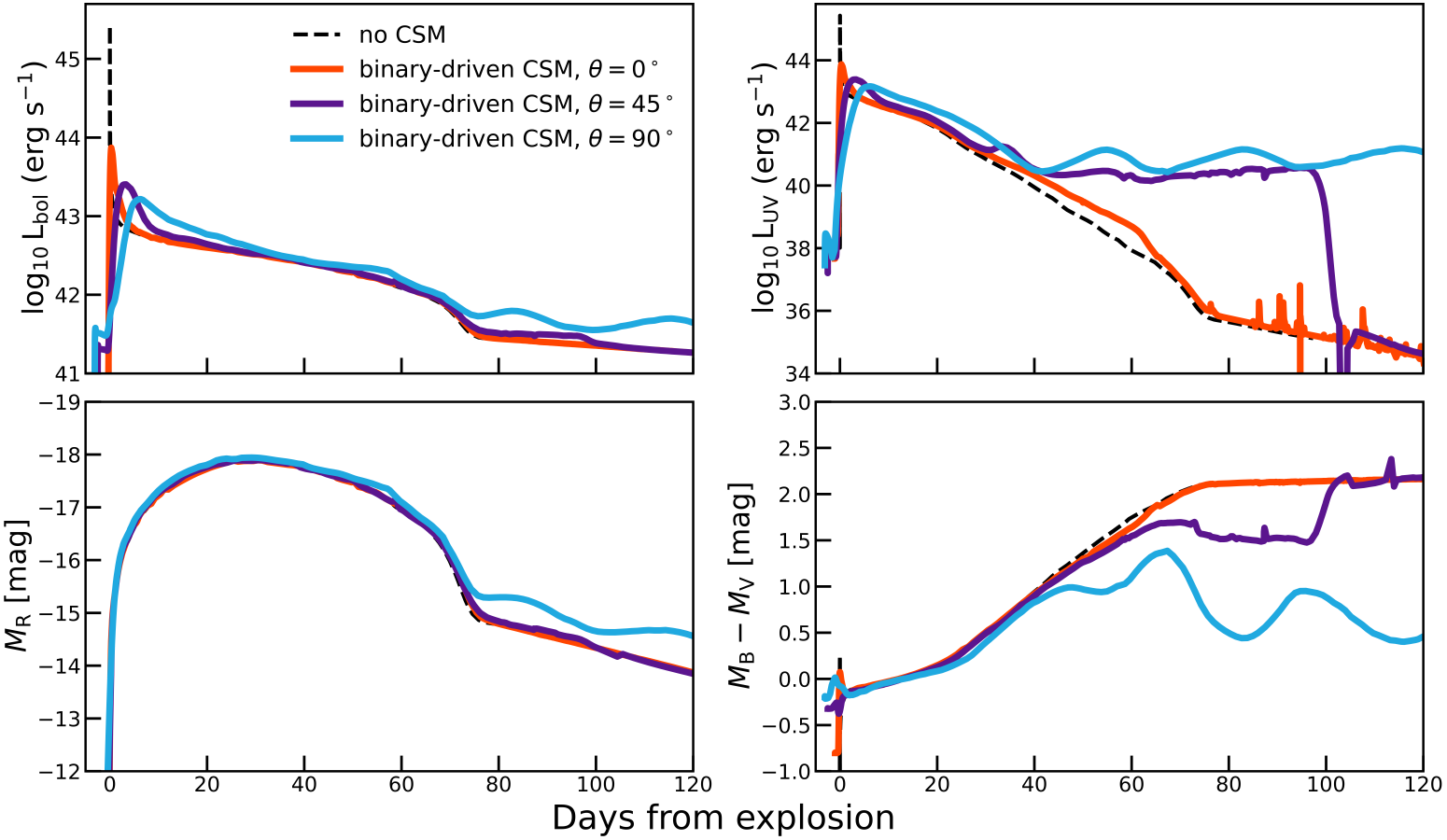}
\vspace{-3mm}
\caption{Bolometric (\textit{top left}), UV ($\lambda\leq320$ nm, \textit{top right}), and $R$-band (\textit{bottom left}) light curves from the q55\_p26 model, interacting with binary generated CSM along different lines of sight. The B-V color evolution of these models is also shown (\textit{bottom right}). Luminosity due to ejecta-CSM interaction increases with increased CSM mass along the line of sight (maximum for the equatorial direction), and is found to reside primarily in the bluer bands. The transient also remains bluer for longer with increased CSM. Additionally, light curves in the equatorial directions exhibit oscillations that originate from ejecta colliding with the undulating CSM seen in Figures.~\ref{fig:csm_slice} and~\ref{fig:csm_1D_slice}.}
\label{fig:lc_5526_csm}
\end{figure*}

At early times ($\lesssim10$ days), the models exhibit the strongest angular dependence. Photospheric radii and velocities along the equatorial direction ($\theta\approx90^{\circ}$) remain systematically larger than those in the polar or oblique directions. This suggests that the photosphere remains embedded within rapidly expanding ejecta in the early days, even along the equatorial direction. Indeed, we find the $\tau=2/3$ surface to reside in the ejecta for all models. Thus, despite the presence of substantial CSM, the photosphere does not reside within a slow, optically thick unshocked CSM shell, unlike the canonical picture for strongly interacting Type IIn SNe \citep{Chugai+2004MNRAS,Smith+2008ApJ,Moriya+2013MNRAS}. It is worth noting that the inferred mass-loss rates in these SNe often exceed the mass-loss rates in our models by several orders of magnitude ($\sim10^{-2}-1\,M_{\odot}\,\mathrm{yr}^{-1}$). Instead, the primary effect of the equatorially enhanced CSM in these models is to modify the optical depth and thermalization structure of the outer ejecta, delaying the recession of the photosphere into deeper and slower layers along viewing directions with the largest CSM column density. The higher photospheric temperatures along polar directions (where one expects lowest CSM column densities) further indicate that geometry-dependent diffusion effects dominate over direct interaction heating during the earliest phases. It is also interesting to note that the photospheric properties in the polar direction are quite similar to the case with no CSM interaction (black squares).

These earliest epochs are also where the limitations of our method are likely most severe. In reality, both the radiation field and the shock interaction itself are multidimensional. Multi-dimensional radiation transport and non-radial shock geometry \citep{Vlasis+2016MNRAS,McDowell+2018ApJ,Suzuki+2019ApJ} can redistribute energy in ways absent from our independent radial-column treatment. However, after the first $\sim$\,10 days, the photospheric properties converge substantially across viewing angles. The radius and velocity evolution become broadly similar, while the temperature differences steadily diminish and converge near the epoch of hydrogen recombination ($T\mathrm{_{ph}}\sim6000$\,K) at around $30$ days. This suggests that the photosphere rapidly recedes out of the asymmetric outer interaction region and becomes increasingly dominated by the homologously expanding ejecta. We therefore expect the direction-dependent 1D approximation to provide a reasonable description of the qualitative angle-dependent light curve evolution beyond the earliest phases ($\sim$\,10 days).

\subsection{Light curves in the presence of CSM}
\label{subsec:light_curves_with_csm}

In \autoref{fig:lc_5526_csm}, we present the bolometric, UV ($\lambda\leq320$ nm), and $R$-band light curves, together with the $B-V$ color evolution, for the q55\_p26 model viewed along different lines of sight. The presence of equatorially-enhanced CSM introduces strong viewing-angle dependence, particularly at blue wavelengths and at late times, while the redder optical bands remain comparatively insensitive to orientation.

\begin{figure*}
\centering
\includegraphics[width=0.98\textwidth]{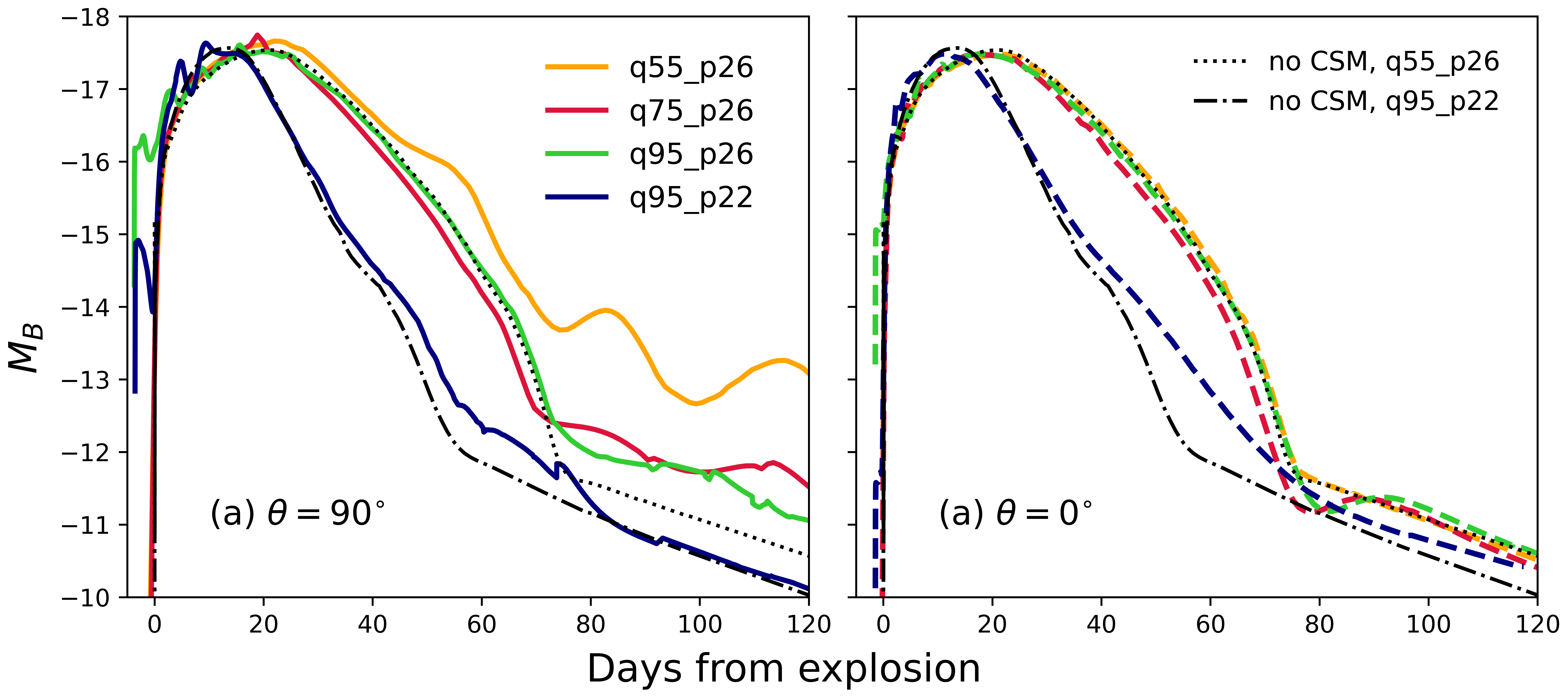}
\vspace{-2mm}
\caption{$B$-band light curves in the equatorial (\textit{left panel}, solid curves) and polar directions (\textit{right panel}, dashed curves) from all models with significant terminal mass loss. Light curves from the q55\_p26 and q95\_p22 models, assuming no CSM, are included for ease of comparison. Binary interaction strongly alters the envelope mass and terminal mass-loss rate from the progenitor (identical at ZAMS for all of these models), and subsequently the SN light curves. As expected from the geometry of the resulting CSM, light curves in the equatorial directions are affected the most. The dominant governing factor for the light curves in the polar direction is the final envelope mass of the progenitor (see Table~\ref{tab:mesa_models}).}
\label{fig:lc_all_csm}
\end{figure*}

The bolometric light curves (top left panel) show seemingly modest angular variation, with the most significant deviations due to viewing angle occuring at the early epochs ($\lesssim20$ days) and at late epochs ($\gtrsim20$ days). These models are also found to suppress the sharp hour-scale shock breakout peak present in the light curve with no CSM interaction (black dashed lines). The sharp peak is broadened over a few days with a decrease in luminosity, with the effect strengthening as more CSM appears in the line of sight. This is a clear sign of radiation being reprocessed by optically thick material, as found by previous studies \citep{Suzuki+2019ApJ,Khatami+2024ApJ}. The light curve viewed from the equatorial direction ($\theta=90^\circ$), corresponding to the densest CSM column, remains systematically brighter after $\sim60$ days and develops a broad late-time plateau absent in the polar views. This reflects sustained ejecta-CSM interaction and kinetic to thermal energy conversion in equatorial directions.

Examining the UV (top right panel) and the $R$-band (bottom left panel) light curves, we find that most of the interaction-powered luminosity resides in the bluer bands. In both cases, all models evolve similarly during the first $\sim$\,40 days, indicating that the early emission is dominated by the expanding ejecta. At later epochs, bluer band light curves observed from the equatorial or oblique directions decline much more slowly compared to their polar counterparts, while the $R$-band evolution remains comparatively unaffected. This implies that the SN stays bluer when more CSM is encountered along the line of sight, as reinforced by the $B-V$ color evolution (bottom right panel) and as seen in previous studies \citep{Moriya+2018MNRAS,Dessart+2022AandA}. The $U$-band light curves (not shown here) also show signs of interaction similar to the UV light curves.

The differing late-time morphologies further highlight the extent to which viewing angle effects shape the observed phenomenology. The equatorial light curve, particularly in UV or bluer optical bands such as $U$, develops pronounced $\sim$\,30 day undulations, qualitatively reminiscent of the light curve variability observed in some interacting transients such as SN 2022jli \citep{Moore+2023ApJ}, SN 2022esa \citep{Maeda+2026PASJ} or SN 2024hpj \citep{Salmaso+2026AandA}. In our models, these fluctuations arise naturally from the spiral density structure of the L2 outflow, causing the interaction region and photosphere to encounter alternating overdense and underdense regions with time. The light curves viewed from polar or oblique directions ($\theta=0^\circ\mathrm{\,or\,}45^\circ$), on the other hand, show weaker and less persistent signs of interaction, resembling more moderately interacting Type IIn transition events such as SN 1998S \citep{Fassia+2000MNRAS}, SN 2013by \citep{Valenti+2015MNRAS} or rapidly declining Type IIL events. While the present work does not argue for a universal binary origin for all of these transients, the models demonstrate that geometry alone can shift the same explosion into very different regions of observed SN phenomenology space. We return to this point in \autoref{sec:discussion}.

In addition to geometry, the light curve from the explosion of a star can be strongly influenced by the properties of its companion. This is demonstrated by \autoref{fig:lc_all_csm}, where we plot the $B$-band light curves from all our models with significant pre-explosion mass loss (models q**\_p26 and q95\_p22), along the equatorial (left panel) and polar (right panel) directions. The effects of ejecta-CSM interaction are more prominent for the equatorial light curves. By comparing against light curves for the q55\_p26 model (dotted lines) and the q95\_p22 model assuming no CSM\footnote{The q75\_p26 and q95\_p26 models with no CSM have light \\curves similar to the corresponding q55\_p26 light curve, as seen in \autoref{fig:lc_no_csm}, and are thus not included in \autoref{fig:lc_all_csm}.}, we see that circumstellar interaction more strongly affects the light curve at late times. As expected, the degree of luminosity enhancement due to circumstellar interaction varies directly with the terminal mass-loss rate, with the least luminosity enhancement for the shorter period model (q95\_p22). This model exhibits the lowest pre-core collapse mass-loss rates among all models experiencing Case C RLOF (Table~\ref{tab:mesa_models}).

By contrast, the polar light curves are less sensitive to the terminal mass-loss rate. Here, the dominant factor in shaping the light curves is the envelope mass of the progenitor at core collapse, as seen in \autoref{fig:lc_no_csm} (in the no CSM case) and as found by \cite{Dessart+2024AandA}. Models with the largest initial orbital period ($P_i=2600$ days) retain the most envelope mass (Table~\ref{tab:mesa_models}), leading to sustained luminosity over $\sim$\,40 days. Systems with tighter orbits (such as q95\_p22) retain less envelope mass and thus the SN luminosity starts declining after $\sim$\,20 days.

\begin{figure*}
\centering
\includegraphics[width=0.99\textwidth]{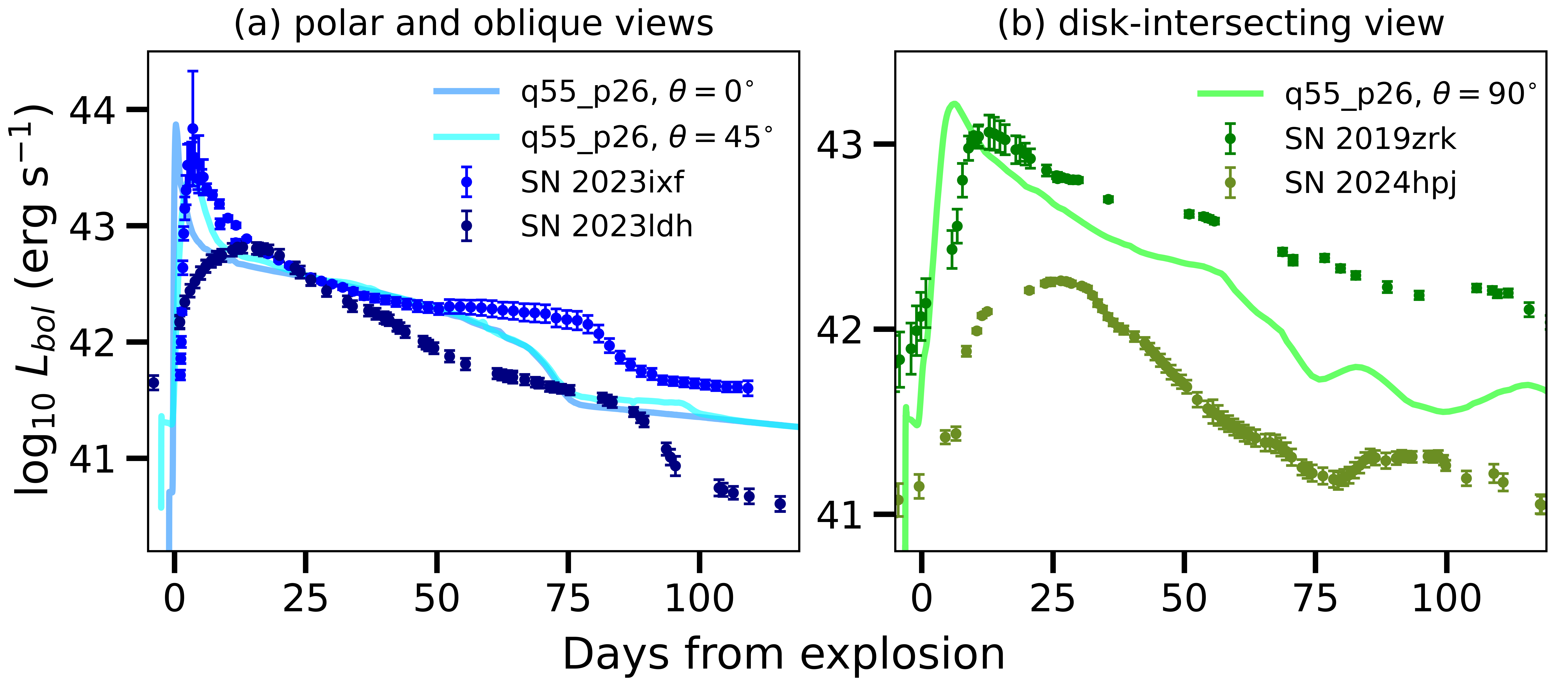}
\vspace{-2mm}
\caption{Comparison of light curves from the q55\_p26 models with some observed interacting transients. \textit{Left}: Light curves viewed along polar ($\theta=0^{\circ}$) and oblique ($\theta=45^{\circ}$) directions exhibit short-lived CSM interactions and a sharp decline after $\sim80-100$ days, similar to SNe 2023ixf and 2023ldh. \textit{Right}: By contrast, light curves viewed along the equatorial ($\theta=90^{\circ}$) direction exhibit sustained CSM interaction due to large CSM column density, along with distinct undulations in the light curve due to interaction with the spiral structures seen in \autoref{fig:csm_slice}. These features are also shown by transients such as SNe 2019zrk or 2024hpj (right panel).}
\label{fig:comparisons_to_data}
\end{figure*}

\section{Discussion} \label{sec:discussion}

\subsection{Diversity of SNe arising from binary systems}

Ejecta-CSM interaction has been recognized to be fairly common in CCSNe \citep{Fraser2020RSOS,Bruch+2023ApJ}, and is found to span a diverse continuum of CSM properties such as density, extent, and geometry \citep{Smith2017hsn..book,Ransome+2025ApJ}. As shown in Section~\ref{subsec:light_curves_with_csm}, substantial diversity in SN light curves and inferred CSM properties can be produced by aspherical CSM interaction, even for the same underlying explosion. This diversity is particularly striking given the relatively modest region of parameter space explored here, and suggests that at least part of the observed spread among Type II SNe may arise from CSM interaction geometry rather than intrinsically distinct explosions. Here, we explore this possibility by comparing our synthetic light curves to observed interacting CCSNe.

We compare the synthetic light curves from the q55\_p26 models along three observer orientations ($\theta=0^\circ,\,45^\circ,\,\mathrm{and}\,90^\circ$) with a sample of interacting transients (\autoref{fig:comparisons_to_data}). The comparison sample consists of pseudo-bolometric light curves assembled from the literature, including the nearby SN 2023ixf \citep{Martinez2024AandA,Hsu_2025,Li+2025AandA}, which exhibits properties intermediate between the Type IIP and IIL subclasses, and the Type IIn SNe 2019zrk \citep{Fransson+2022AandA}, 2023ldh \citep{Pastorello2025AandA}, and 2024hpj \citep{Salmaso+2026AandA}. For all light curves but that for SN 2023ixf, we use bolometric reconstructions following the procedure described in \cite{Salmaso+2026AandA}.

The polar and oblique viewing angles ($\theta=0^\circ$ and $45^\circ$) intersect a relatively small amount of CSM. As such, they produce light curves characterized by early CSM interaction followed by a relatively rapid decline after the plateau phase, resembling events such as SNe 2023ixf and 2023ldh. By contrast, the equatorial viewing angle ($\theta=90^\circ$), where the line of sight intersects the densest CSM, produces a more prolonged interaction-powered decline with late-time undulations, qualitatively similar to transients such as SN 2019zrk and SN 2024hpj. Notably, SNe 2019zrk, 2023ldh and 2024hpj all exhibited similar eruptive activity prior to explosion \citep{Fransson+2022AandA,Pastorello2025AandA}, reminiscent of the Type IIn SN 2009ip \citep{Pastorello+2013ApJ}. Such pre-explosion eruptions have indeed been proposed to arise from binary interaction and mergers, e.g., SN 2022mop \citep{Brennan+2025arXiv}. Despite this similarity, the late-time light curves of these transients differ substantially, analogous to the diversity produced by different viewing angles in our models. Our models do not reproduce the absolute luminosities or decline times of these events, particularly evident for SNe 2023ldh and 2024hpj, presumably due to differences in explosion energies and ejecta masses between the events and our models. Nevertheless, these comparisons demonstrate that variations in CSM geometry and viewing angle can produce light curves spanning a broad range of observational classes, even for an otherwise identical explosion.

\subsection{Fitting to model light curves}
\label{subsec:fitting}

The qualitative comparisons in the previous section suggest that a single CCSN in a binary system can resemble different SN subclasses, depending on the viewing angle. We now ask whether such apparently different transients, if observed, would also yield different inferred progenitor properties. We therefore fit a subsample of our synthetic light curves using a grid of 228016 Type II SN models from \cite{Moriya+2023PASJ}. In this subsample, we only consider systems with a strong departure from spherical symmetry, that is, only the systems with significant mass loss immediately before core collapse (models q**\_p26 and q95\_p22; see Table~\ref{tab:mesa_models}).

We adopt the model bolometric light curves for the fitting procedure, motivated by observational studies where fitting is typically performed on the bolometric light curve interpolated from data available in UVOIR filters. We also perform fits to data at individual filters and obtain similar qualitative results. As discussed in Section~\ref{subsec:photosphere_ev_csm}, we expect multi-dimensional effects to be the most prominent during the early phase ($\approx10$ days), which are not accounted for in this study. We therefore exclude the early phase from our fits. It is important to note that this phase is also affected by shock cooling effects, and can be analyzed to infer the radius and mass of the cooling envelope \citep{Piro21,SapirWaxman23}. We defer such analysis for a future multi-dimensional modeling study, and focus here on the light curves during $t\gtrsim5-10$ days). 

\begin{figure}
    \centering
    \includegraphics[width=8 cm, height=6cm]{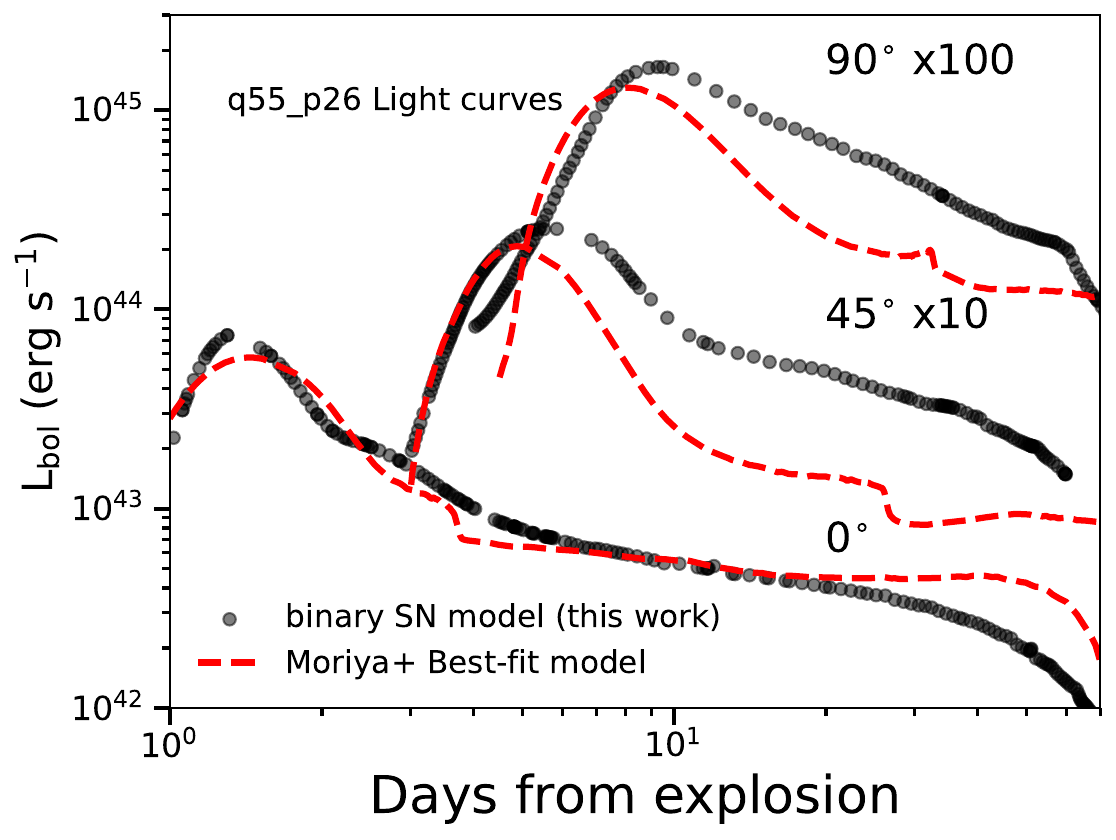}
    \caption{A comparison of the best-fit model from the grid of Type II SN light curves in \cite{Moriya+2023PASJ} to the q55\_p26 bolometric light curves at viewing angles $\theta=0^\circ,45^\circ,\mathrm{and\,}90^\circ$. For easier visualisation, we adjust the bolometric luminosities of the latter two  by factors of 10 and 100, respectively. }
    \label{fig:bestfit_typeII}
\end{figure}

We also opt to fit the light curve until $\approx80$ days to simulate typical observations, where a transient can only be observed for a finite duration before its brightness drops below detection thresholds. We choose 80 days as the cutoff as for an object at 50 Mpc (a relatively average distance), the B/V band magnitudes would fall below 20th magnitude and thus not be detectable with current survey capabilities.  We run fits by iterating against the entire grid of models and select the model that minimizes $\chi^2$ as the best fit (while also capturing the rise and fall of the lightcurve--we exclude fits that are due to late-time similarity in the lightcurves but do not match the rise or decline time). We compute the CSM mass from the model grid as well, noting that qualitatively similar results are obtained when surrogate models \citep[from][]{Sarin24} are fit using Redback \citep{Sarin24}. Examples of the fits (for the q55\_p26 models) are shown in \autoref{fig:bestfit_typeII}.

\begin{deluxetable*}{ccccccccc}
\tablecaption{Parameters derived from fitting the grid of type II progenitor models \citep{Moriya+2023PASJ} to the light curves computed in this study.}
\label{tab:inferences}
\tablehead{
\colhead{Model} & \colhead{Viewing angle} & \colhead{$M_{\mathrm{ZAMS}}$} & \colhead{$E_{\mathrm{exp}}$} & \colhead{$M_{\mathrm{Ni}}$} & \colhead{$\mathrm{log}_{10}\left(\dot{M}_{\mathrm{CSM}}\right)$} & \colhead{$R_{\mathrm{CSM}}$} &\colhead{$\tilde\beta$}  & \colhead{$M\mathrm{_{CSM}}$} \\[-1ex]
\colhead{Name} & \colhead{(degrees)} & \colhead{($M_{\odot}$)} & \colhead{($10^{51}$ ergs)} & \colhead{($M_{\odot}$)} & \colhead{($M_{\odot}$/year)} & \colhead{($10^{14}$ cm)} & \colhead{} & \colhead{($M\mathrm{_{\odot}}$)} \\
\colhead{(1)} & \colhead{(2)} & \colhead{(3)} & \colhead{(4)} & \colhead{(5)} & \colhead{(6)}  & \colhead{(7)} & \colhead{(8)} &\colhead{(9)}\\[-2ex]
}
\startdata
%\hline
%\multicolumn{9}{l}{\textit{$\chi^2$-based fits}} \\
%\hline
\multirow{3}{*}{q55\_p26} & 90 &14&  0.5&  0.001&  -2.5&  10&  1& 0.15\\
 & 45 &  10&  0.5&  0.001&  -3.0&  10&  3 &0.19\\
 &  0 &10&  3.5& 0.04&  -3.5&  4&  1&0.009\\
\hline
\multirow{3}{*}{q75\_p26} & 90 & 16 & 1.0& 0.001 &-2.5& 10& 0.5&0.12\\
 & 45 & 16 & 1.0& 0.001&-2.5 & 10& 1.5&0.22\\
 &  0 & 10 & 3.0 & 0.01 & -3.5 & 4 & 1.5 &0.02 \\
\hline
\multirow{3}{*}{q95\_p26} & 90 & 10 & 1.0 & 0.01 & -2.5 & 10& 0.5 &0.11\\
 & 45 & 10 & 1.0 & 0.001 & -2.5 & 10 & 1 &0.13\\
 &  0 & 10 & 2.5 &0.01 & -3.5 & 6 & 2&0.03 \\
\hline
\multirow{3}{*}{q95\_p22} & 90 & 10 & 0.5 & 0.001& -3 &10 & 4&0.34 \\
 & 45 & 10 & 0.5 & 0.001 & -3.5 & 10 & 2.5&0.06 \\
 &  0 & 12 & 3& 0.2 & -3.5& 6 & 1.5&0.02 \\
 \hline
 \enddata
\tablenotetext{}{The $\chi^2$-based fits to our model light curves (at three separate viewing angles), computed using \cite{Moriya+2023PASJ}'s grid of Type II SN models. Columns are as follows: (1) model name, (2) observer viewing angle, and (3)-(9) parameters inferred from the fits. We note that these parameters are not necessarily the same as the model intrinsic parameters mentioned in Table~\ref{tab:mesa_models}. The inferred parameters are: (3) ZAMS mass of the progenitor, (4) explosion energy, (5) $^{56}$Ni mass, (6) terminal mass loss rate from the progenitor, (7) outer radius of the CSM, (8) wind acceleration parameter, and (9) total CSM mass \citep[see][for details]{Moriya+2023PASJ}. Note that the ZAMS mass, explosion energy, and $^{56}$Ni mass for all our models are set to $16M_{\odot}$, $10^{51}$ ergs, and $0.04M_{\odot}$), respectively (see Section~\ref{subsubsec:SN}). The true mass-loss rates for the models q55\_p26, q75\_p26, q95\_p26, and q95\_p22 are $\sim10^{-3},\,10^{-4},\,10^{-4},\,\mathrm{and}\,10^{-5}\,M_{\odot}\,\mathrm{yr}^{-1}$, respectively (see Table~\ref{tab:mesa_models}).}
\end{deluxetable*}

 The parameters encompassed by the models in the grid are the progenitor ZAMS mass ($M_{\mathrm{ZAMS}}$), terminal mass-loss rate of the progenitor ($\dot{M}_{\mathrm{CSM}}$), the outermost extent of the CSM (or CSM radius; $R_{\mathrm{CSM}}$), explosion energy ($E_{\mathrm{exp}}$), $^{56}$Ni
mass ($M_{\mathrm{Ni}}$), and the wind acceleration parameter $\tilde{\beta}$. The last parameter\footnote{\cite{Moriya+2023PASJ} refer to this parameter as $\beta$; we adopt \\ $\tilde{\beta}$ instead since $\beta$ is used to denote the mass transfer \\ efficiency in our models.} controls the velocity profile (and hence the density profile) of the wind-like CSM in these models. For high values of $\tilde{\beta}\,(\gtrsim3)$, the CSM density profile is relatively steeper than $1/r^2$ \citep[see Figure 1 of][]{Moriya+2023PASJ}. For lower $\tilde{\beta}$, the CSM density profile immediately drops to that of a steady stellar wind. Given lower densities and steeper density profiles away from the equatorial plane in our hydrodynamic models (\autoref{fig:csm_1D_slice}), we therefore expect smaller inferred $\dot{M}_{\mathrm{CSM}}$ and larger $\tilde{\beta}$ as the observing angle $\theta$ (from the binary axis) becomes smaller. We also expect inferred values of $R_{\mathrm{CSM}}$ to be smaller in the polar or oblique viewing directions. Finally, we note that the progenitor ZAMS mass (equal to $16M_{\odot}$
for all models in this study) is likely to be underpredicted by these fits. This is expected since the fitting procedure consists of parameter estimation via comparison against a grid of isolated stellar models that do not account for binary-driven mass loss.

%The early phases of the bolometric light curve are affected by shock breakout and shock cooling effects not captured by the models, and thus we opt only to fit once the supernova has begun its rise to bolometric peak (at $\sim >$ 5 days for the $45/90^{\circ}$ cases, and $\sim >$ 1 day for the 0 degree case). 

%We adjusted the range over which we fit to ensure a good quality of fit by eye (i.e., in certain cases fitting over a certain time range caused very obviously poor fits, in which case we adjusted the range).

In Table~\ref{tab:inferences}, we list the parameters inferred from our binary model light curves. We also provide the total CSM mass, calculated using the CSM density grid in \cite{Moriya_2024} and the inferred CSM parameters ($\dot{M}_{\mathrm{CSM}}$, $R_{\mathrm{CSM}}$, and $\tilde{\beta}$). The expected trend emerges in these fits: the amount of CSM inferred is largest for the equatorial view. Inferred values of $R_{\mathrm{CSM}}$ and $M_{\mathrm{ZAMS}}$ are also as expected:  $M_{\mathrm{ZAMS}}$ is usually under-predicted as per expectations, with major exceptions for the q75\_p26 models along $\theta=45^{\circ}$ and $90^{\circ}$, where the inference is surprisingly accurate. The reason for this exception remains unclear. The inferred mass-loss rates ($\dot{M}_{\mathrm{CSM}}$) reveal an interesting property of the fits to our models. Looking at \autoref{fig:csm_1D_slice}, one would expect inferred mass-loss rates of $\approx10^{-3}M_{\odot}\,\mathrm{yr}^{-1}$ for e.g., the q55\_p26 model in the equatorial direction, and somewhat lower mass-loss rates in the oblique and polar directions. Instead, we find that the fits strongly over-predict the mass-loss rate from the equatorial light curves for all models, although the mass-loss rates inferred for the oblique and polar directions are progressively lower as per the expected trend. On the other hand, considerable spread is found in the $\tilde{\beta}$ parameter over all models, without a clear emerging trend. We also note that the explosion energy (equal to $10^{51}$ ergs for all models) is systematically over-predicted as the viewing direction moves away from the equatorial plane to the binary axis. The total CSM mass ($M_{\mathrm{CSM}}$) inferred is much more in the equatorial or oblique directions compared to the polar direction, although it is not clear why the oblique direction ($\theta=45^{\circ}$) yields the maximum amount of CSM. We also note that $M_{\mathrm{CSM}}$ is expected to be much less than the total mass lost from the binary system (as reported in Table~\ref{tab:mesa_models}), since $M_{\mathrm{CSM}}$ is inferred here based on ejecta-CSM interaction for $\approx80$ days and accounts for mass loss only in the final $\sim1000$ days prior to core collapse of the donor. The total mass lost from the binary, on the other hand, takes into account mass loss that spans over a period of $\sim10^4$ years for the Case C RLOF systems considered here.

In summary, we find that binary-driven envelope stripping and equatorially-confined CSM can introduce significant errors in the interpretation and analysis of SN light curves, particularly in cases where inferences involve assumptions of isolated stellar evolution and steady-state (either constant speed or accelerated) stellar winds. In particular, inferences for the ZAMS mass, explosion energy, and terminal mass-loss rates are found to be the most sensitive to this phenomenon.

\section{Conclusion} \label{sec:conclusion}

In this study, we evolve stellar binary models using \mesa up to the core collapse of the initially more massive star, accounting for non-conservative mass transfer and mass loss. Motivated by recent theoretical studies \citep{Lu+2023MNRAS,Scherbak+2025ApJ}, gas lost from the binary system is modeled using the 3D expanding mesh hydrodynamics code \sprout as an outflow from the outer Lagrange (L2) point. We find that the large specific angular momentum of the outflow, combined with the gravity of the binary, generates dense, equatorially-enhanced CSM. Using hydrodynamic profiles sampled along different lines of sight through this aspherical CSM, we model the ensuing SN of the mass donor star using the 1D radiation hydrodynamics code \texttt{Stella} to obtain direction-dependent light curves. We find that CSM asphericity and the resulting viewing-angle effects significantly influence the light curves of our models, particularly in the bluer bands and at late times. Our principal novel findings are summarized below:

\begin{enumerate}
    \item Stable late-stage (Case C) RLOF in massive binaries naturally generates dense, equatorially-concentrated CSM through non-conservative L2 mass loss. Systems with lower mass companions exhibit the strongest mass loss (up to $10^{-3}M_{\odot}\mathrm{\,yr^{-1}}$ in our models). Explosion of the donor in such systems also therefore exhibit the most significant viewing angle effects.

    \item The resulting CSM exhibits strong angular variation for both density and velocity. While equatorial sight-lines are broadly consistent with steady wind-like profiles ($\rho\propto r^{-2}$), polar directions can exhibit substantially steeper effective density profiles (up to $\rho\propto r^{-4}$), despite arising from the same underlying outflow.

    \item In the presence of binary-generated CSM, the same SN explosion can appear as qualitatively different interacting transients in different viewing directions. Amongst our models, the contrast between the bolometric luminosity at late times ($\sim100$ days) for the same SN viewed in different directions can reach a factor of $\approx3$.

    \item The strongest signatures of interaction occur for equatorial viewing directions, where enhanced ejecta-CSM interaction stays bluer for longer durations and generates characteristic light curve undulations arising from the spiral structure of the L2 outflow. By contrast, polar viewing directions remain comparatively weakly affected by CSM interaction. For instance, the UV ($\lambda\leq320$ nm) luminosity for the q55\_p26 model along the equatorial direction exceeds that for the same model but in the polar direction by a factor of $6\times10^5$ at 100 days post explosion.

    \item We compare our models to various SNe belonging to different subclasses of  interacting Type II SNe, such as SN 2023ixf and SN 2019zrk. These comparisons suggest that at least part of the observed diversity among CCSNe may arise from asymmetric binary-generated CSM and orientation effects, rather than requiring intrinsically different explosions or eruptive mass-loss histories.

    \item Binary interaction can introduce significant errors in inferences drawn from observed Type II SNe using state-of-the-art 1D SN inference frameworks. In particular, we find that the inferred progenitor ZAMS mass, explosion energy, and $^{56}$Ni mass can have errors of about $50\%$. We further show that such frameworks can infer CSM masses varying by a factor of a few tens for the same explosion, but observed along different viewing angles.

\end{enumerate}

Our results highlight the need for caution when interpreting CCSNe using spherically symmetric models. Although the binary parameter space explored here represents only a subset of possible evolutionary pathways, binary population synthesis studies suggest that late-stage mass transfer in wide RSG binaries may produce interacting explosions at a rate of order $\approx5\%$ of all CCSNe \citep{Ercolino+2024AandA,Ercolino+2026AandA}. This fraction is comparable to the observed rate of strongly interacting CCSNe \citep[$\approx10\%$;][]{Perley+2020ApJ}, suggesting that binary interaction may represent an important pathway for producing interacting CCSNe. Yet, a number of important uncertainties remain. The present study explores only a limited region of the binary parameter space, considering a single primary ZAMS mass at solar metallicity and a small set of initial mass ratios and orbital periods. Extending this work to a wider range of binary properties will be essential to map the diversity of CCSNe in binaries, as the efficiency of binary mass transfer, the degree of envelope stripping, and the resulting CSM morphology are all expected to vary across the parameter space. Such studies should also investigate sub- and super-solar metallicities, motivated by the observed dependence of CCSN populations on host galaxy metallicity \citep{Modjaz+2011ApJ,Anderson+2015PASA,Pessi+2023AandA,Ganss+2025MNRAS}. The hydrodynamic calculations presented here neglect physical processes such as radiative cooling and magnetic fields that may influence the morphology and density contrast of the outflow \citep{Scherbak+2026PASP}. Furthermore, our observable predictions are derived from direction-dependent 1D radiation-hydrodynamic calculations. Although this approach captures the large-scale asymmetry of the binary-generated environment, it does not account for fully multidimensional ejecta--CSM interaction. Future work incorporating a broader range of binary evolutionary pathways together with multidimensional radiation-hydrodynamic modeling will therefore be necessary to quantify the full observational consequences of asymmetric binary-generated CSM and to determine the extent to which viewing-angle effects contribute to the observed diversity of CCSNe.

Nevertheless, the results presented here demonstrate that binary interaction can generate strongly aspherical circumstellar environments capable of producing substantial photometric diversity from a common underlying explosion. Incorporating asymmetric binary-generated CSM into future inference frameworks will therefore be important for robust interpretation of interacting transients. Since the strongest signatures of asymmetric ejecta-CSM interaction emerge predominantly at late times and in blue/UV wavelengths, continued late-time monitoring by upcoming wide field-of-view UV observatories, such as ULTRASAT \citep{Shvartzvald+2024ApJ}, will be especially important for constraining the true structure and geometry of the CSM surrounding CCSNe.

\acknowledgments
 
We thank Jared Goldberg and other members of the \mesa team, along with Sergei Blinnikov, for their copious helpful inputs with the numerical setup. We also thank Takashi Moriya for kindly sharing the grid of models presented in \cite{Moriya+2023PASJ}, as well as for providing helpful insights on the manuscript. I.S. acknowledges financial support from the SOXS project. Numerical calculations were performed on the Rivanna computing cluster at University of Virginia.

\software{\mesa \citep{Paxton2011, Paxton2013, Paxton2015, Paxton2018, Paxton2019, Jermyn2023}, 
          \sprout\, \citep{Mandal+2023_sprout},
          \stella \citep{Blinnikov+1993AandA,Blinnikov+1998ApJ,Blinnikov+2004ApandSS,Blinnikov+2006AandA},
          NumPy \citep{numpy},
          Matplotlib \citep{matplotlib}.
}

\vspace{15mm}

\bibliographystyle{apj} 
\typeout{}
\bibliography{smbib}

\end{document}